\documentclass[article]{elsarticle}

\usepackage{lineno,hyperref}
\usepackage{cite}
\usepackage{amsmath,amssymb,amsfonts}
\usepackage{algorithmic}
\usepackage{graphicx}
\usepackage{textcomp}
\usepackage{hyperref}
\usepackage{booktabs}
\usepackage{multirow}
\usepackage{colortbl} 
\usepackage[table,xcdraw]{xcolor} 
\usepackage{fancyhdr} 

\journal{Neurocomputing}

\begin{document}
\let\WriteBookmarks\relax
\def\floatpagepagefraction{1}
\def\textpagefraction{.001}

\begin{frontmatter}

\title{Embedding Similarity Guided License Plate Super Resolution}

\address[first_address]{CNRS, Univ. Poitiers, XLIM, UMR 7252, France}

\author[first_address]{Abderrezzaq Sendjasni*}
\author[first_address]{Mohamed-Chaker Larabi}

\cortext[1]{Corresponding author}

\begin{abstract}

Super-resolution (SR) techniques play a pivotal role in enhancing the quality of low-resolution images, particularly for applications such as security and surveillance, where accurate license plate recognition is crucial. This study proposes a novel framework that combines pixel-based loss with embedding similarity learning to address the unique challenges of license plate super-resolution (LPSR). The introduced pixel and embedding consistency loss (PECL) integrates a Siamese network and applies contrastive loss to force embedding similarities to improve perceptual and structural fidelity. By effectively balancing pixel-wise accuracy with embedding-level consistency, the framework achieves superior alignment of fine-grained features between high-resolution (HR) and super-resolved (SR) license plates. Extensive experiments on the CCPD and PKU dataset validate the efficacy of the proposed framework, demonstrating consistent improvements over state-of-the-art methods in terms of PSNR, SSIM, LPIPS, and optical character recognition (OCR) accuracy. These results highlight the potential of embedding similarity learning to advance both perceptual quality and task-specific performance in extreme super-resolution scenarios. 
\end{abstract}


\end{frontmatter}
\section{Introduction}
\label{sec:intro}

Single image super-resolution (SISR) is a well-known research field in computer vision focused on enhancing spatial resolution and visual fidelity of low-resolution images. Its significance lies in the ability to reconstruct high-resolution details from degraded visual data, thereby improving image quality across diverse applications, including digital photography~\citep{wang2018esrgan, Ledig_2017_CVPR}, medical imaging~\citep{van2012super}, and video surveillance~\citep{9663831}. In particular, SISR has been increasingly applied to enhance license plate (LP) images, where the clarity and legibility of such critical visual data are paramount for effective and reliable automated recognition systems. 

\begin{figure}[h]
     \centering
     \includegraphics[width=\textwidth]{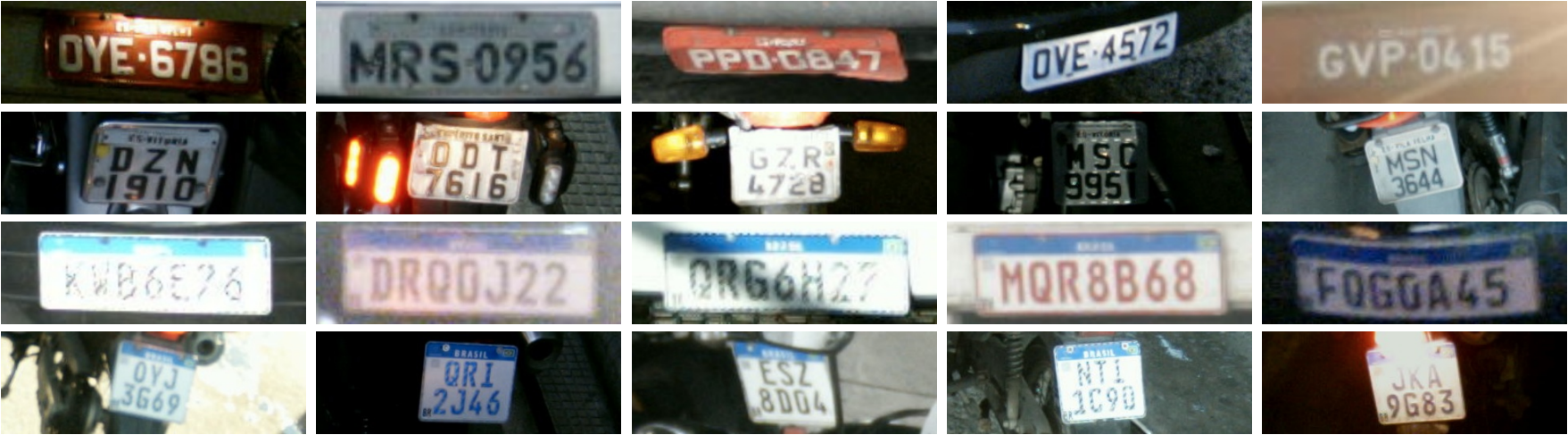}
    \caption{Example of license plates taken under different conditions, showcasing difficulties to properly read the plates in some cases~\citep{laroca2022cross}.}
    \label{fig:ex_lps}
 \end{figure}

License plate recognition (LPR) systems are integral components of modern surveillance, traffic management, and security applications~\citep{6339122}. However, the efficacy of these systems heavily relies on the quality of the captured images~\citep{NASCIMENTO202369}. In real-world scenarios, LP images captured by surveillance cameras or other sources often suffer from visual degradations, such as low resolution, motion blur, and noise. As illustrated in Fig.~\ref{fig:ex_lps}, these issues make it challenging to accurately read the LPs, posing significant difficulties for LPR systems and compromising their accuracy and reliability. Among these challenges, the most critical is the distance at which the images are captured, leading to limited pixel resolution. When images are taken from a great distance, LPs appear smaller within the frame, drastically reducing the number of pixels representing the plate. This loss of resolution causes fine details to become invisible and unrecognizable upon zooming, making it difficult for both LPR systems and human users to interpret and recognize the characters accurately. To address this, advanced image enhancement techniques, such as super-resolution (SISR), are crucial for upscaling low-resolution images and preserving the key details necessary for reliable LPR.

Despite recent advances in the field of SISR~\citep{10041995, 9870558, li2023ntire}, LP super-resolution (LPSR) remains a significant challenge. The unique characteristics of LP images, such as small text, complex backgrounds, varying lighting conditions, and diverse fonts~\citep{gonccalves2019multi, 9959390, 9897178, gong2022unified}, make it difficult for standard SISR models to perform robustly. Traditional SR methods often fail to reconstruct the fine details and sharp edges required to accurately recognizing characters on license plates. Deep learning-based methods, while more robust, often face challenges in striking the delicate balance between enhancing image quality and preserving critical textual information. Achieving high accuracy in LPR systems is crucial, particularly in applications such as law enforcement, where misrecognition or failure to identify a license plate can have serious consequences. This underscores the critical need to develop super-resolution (SR) methods specifically tailored to address these challenges.

Compared to the extensive research on SISR, studies specifically addressing LPSR remain limited. Most existing approaches rely on deep learning, primarily due to its ability to leverage inherent prior knowledge of natural scenes and preserve image details more effectively than traditional methods like interpolation. For example, interpolation-based techniques such as bilinear and bicubic interpolation are simple and fast but often produce blurry images with a loss of fine details. These methods estimate new pixel values based on linear or cubic interpolation of neighboring pixels, resulting in smooth but less detailed outputs. In contrast, deep learning-based approaches, such as convolutional neural networks (CNNs) and generative adversarial networks (GANs), have demonstrated superior performance in LPSR~\citep{pan2024lpsrgan}. Their ability to learn complex patterns enables them to reconstruct high-resolution images with finer details and enhance visual fidelity.

\textcolor{black}{
In this work, we address the limitations of existing LPSR methods by focusing on extreme super-resolution scenarios with a challenging scaling factor of $\times 8$. To tackle the reconstruction of severely degraded license plates with minimal pixel information, we propose a deep learning framework that combines residual dense blocks (RDBs) and channel attention mechanisms to enhance visual quality and textual detail recovery. Our approach integrates pixel-level and embedding-level losses, implemented using a Siamese network to align embeddings of high-resolution (HR) and super-resolution (SR) images. By doing so, the character recovery is improved along with the overall fidelity. Besides, the training strategy based on embedding similarity utilizes contrastive loss~\citep{hadsell2006dimensionality} to minimize the discrepancy between HR and SR embeddings. This strategy preserves fine details that are critical for character recognition. Extensive experiments conducted on both CCPD~\citep{xu2018towards} and PKU~\citep{7752971} datasets, containing real-world license plate images under diverse conditions, demonstrate the effectiveness and robustness of the proposed method. The primary contributions of this work can be summarized as:}

\textcolor{black}{
\begin{itemize} 
    \item We developed a deep learning framework for extreme LPSR with a scaling factor of $\times 8$, leveraging residual dense blocks and channel attention mechanisms to enhance visual quality and recover fine details.
    \item We introduced the pixel and embedding consistency loss (PECL), which integrates pixel-level and embedding-level similarities. A Siamese network and contrastive loss are employed to align and constrain the similarity between HR and SR embeddings.
    \item We conducted a comprehensive evaluation of the proposed method on both CCPD and PKU datasets, demonstrating its robustness and effectiveness across diverse real-world conditions.
\end{itemize}
}

\section{Related work}
\label{sec:related_work}
\subsection{Single image super-resolution}
Single image super-resolution (SISR) has seen extensive research in the past decade, primarily driven by deep learning techniques that aim to reconstruct high-resolution (HR) images from their low-resolution (LR) counterparts. Early methods for SISR were based on interpolation techniques such as bilinear and bicubic interpolation, which, despite their simplicity, often resulted in blurred images with a loss of fine details~\citep{1163711}. These shortcomings prompted the development of more sophisticated methods like sparse coding-based models~\citep{5466111}, and later, deep learning-based approaches such as convolutional neural networks (CNNs) and generative adversarial networks (GANs), which have demonstrated substantial improvements in both perceptual quality and quantitative performance metrics.

CNN-based models like SRCNN~\citep{dong2015image} introduced the concept of end-to-end learning for SISR, laying the groundwork for more complex architectures such as VDSR~\citep{kim2016accurate}, EDSR~\citep{lim2017enhanced}, and RCAN~\citep{zhang2018image}, which leverage residual learning and attention mechanisms to improve the super-resolution performance. GAN-based methods, such as SRGAN~\citep{Ledig_2017_CVPR}, ESRGAN~\citep{wang2018esrgan}, and more recent works like Real-ESRGAN~\citep{wang2021real} and SwinIR~\citep{liang2021swinir}, focus on enhancing the perceptual quality of the super-resolved images by employing adversarial learning and perceptual loss functions based on deep features~\citep{johnson2016perceptual} as well as vision transformers~\citep{dosovitskiy2021an}. These methods have successfully generated SR images with sharper details and more visually appealing results compared to traditional interpolation-based methods.

Despite these advancements, standard SISR models often struggle when applied to domain-specific tasks such as license plate super-resolution (LPSR), where the primary goal is not just to improve image fidelity, but also to preserve critical textual and structural information that is vital for recognition tasks. Indeed, the challenges are more domain-specific and tied to the unique visual characteristics of license plate images, such as small fonts, varying lighting conditions, and complex backgrounds. Traditional SISR models, when directly applied to license plates, tend to fail in recovering the fine-grained details required for character recognition, particularly when dealing with extreme scaling factors such as $\times 8$ or higher.

\subsection{License plate super-resolution}

Several approaches have been proposed to address the previously mentioned challenges by incorporating domain-specific knowledge into the super-resolution pipeline. The work in~\citep{8492768} presented a multi-scale CNN tailored for LPSR, focusing on minimizing the mean squared error (MSE) between HR and super-resolved (SR) license plate images. While effective at enhancing the overall image quality, this approach still struggled with preserving fine textual details, which are critical for accurate LPR. Recent advances in deep learning have paved the way for more sophisticated LPSR models. The work in~\citep{10105735} introduced a GAN-based architecture that incorporates a gradient profile prior to emphasize character boundaries, thus improving the contrast between the characters and the background. Similarly, the authors in~\citep{pan2024lpsrgan} extended the SRGAN framework by adding an optical character recognition (OCR)-based loss function, which directly computes the recognition error between HR and SR images, thus ensuring that the generated SR images retain the legibility of the text. This approach also utilizes perceptual loss functions based on VGG-19~\citep{wang2021real, Simonyan15} to improve both the visual quality and recognition accuracy of the license plates.

Other notable works have explored the use of character-based perceptual losses, where the super-resolution process is guided by intermediate feature representations learned by an OCR network. For example, the work in~\citep{9070241} proposed a loss function based on character classification features, while~\citep{9991753} employed the Levenshtein distance to measure discrepancies between predicted and ground-truth characters. These methods highlight the importance of integrating recognition tasks directly into the loss function, allowing the models to not only enhance the visual quality but also optimize for accurate character recognition.

Despite these advances, several challenges remain in LPSR research. First, many existing approaches are designed for moderate upscaling factors (\textit{i.e.} x2 or x4) and fail to generalize to extreme cases such as $\times 8$ and beyond, where the license plate details are severely degraded. Moreover, current LPSR models often struggle to balance the trade-off between improving perceptual quality and preserving critical textual information. While OCR-guided loss functions have shown promise in mitigating this issue, there is still room for improvement, particularly in cases where the input LR images suffer from extreme distortions, such as motion blur or severe compression artifacts.

This study addresses the limitations of existing LPSR approaches by introducing a novel training strategy that incorporates both pixel-wise and embedding-level losses. By leveraging a combination of perceptual and contrastive loss functions, the proposed method ensures robust super-resolution even at extreme scaling factors, achieving a balance between perceptual quality and the preservation of textual and structural details. The following sections detail the methodology and demonstrate its effectiveness compared to state-of-the-art techniques.

\section{Proposed Methodology}
\label{sec:method}

\subsection{Problem formulation}

The task of SR aims to reconstruct an HR image $\mathbf{I}_{HR} \in \mathbb{R}^{h \times w \times 3}$ from its LR counterpart $\mathbf{I}_{LR} \in \mathbb{R}^{H \times W \times 3}$, where typically $h = s \cdot H$ and $w = s \cdot W$, and $s \in \mathbb{Z}^{+}$ is the upscaling factor. The goal of SR is to recover fine-scale details lost during image degradation processes such as downscaling, compression, or noise corruption. Thus, the problem we try to solve can be formulated as a learning task where the objective is to estimate a function $f_\theta: \mathbb{R}^{H \times W \times 3} \rightarrow \mathbb{R}^{h \times w \times 3}$, parameterized by $\theta$, that maps the LR image $\mathbf{I}_{LR}$ to its super-resolved version $\mathbf{I}_{SR}$:

\begin{equation}
\mathbf{I}_{SR} = f_\theta(\mathbf{I}_{LR}),
\end{equation}

\noindent where the learning task can be framed as an optimization problem with the objective to minimize a composite loss function $\mathcal{L}_{\text{total}}$ over a set of parameters $\theta$, subject to constraints imposed by the nature of the task. Formally, the goal is to find:

\begin{equation}
\theta^* = \arg \min_{\theta} \mathcal{L}_{\text{total}}(\mathbf{I}_{HR}, f_\theta(\mathbf{I}_{LR})),
\end{equation}

\noindent where $f_\theta(\cdot)$ is the parameterized mapping from LR to SR images. The total loss function $\mathcal{L}_{\text{total}}$ is a weighted combination of several distinct loss terms:

\begin{equation}
\mathcal{L}_{\text{total}} = \sum_{i=1}^{N} \lambda_i \mathcal{L}_i,
\end{equation}

\noindent where $\mathcal{L}_i$ represents different loss components corresponding to specific properties or objectives that must be optimized, and $\lambda_i \in \mathbb{R}^{+}$ are scalar weights that control the contribution of each term.

In our method, we model the problem using a patch-based approach. Rather than processing entire HR and LR images, we operate on smaller patches extracted from these images. Formally, let $\mathbf{I}^{HR} \in \mathbb{R}^{C\times H\times W}$ and $\mathbf{I}^{LR} \in \mathbb{R}^{C\times h\times w}$ represent the HR and LR images respectively, where $H > h$ and $W > w$. Instead of directly mapping the whole LR image to its HR counterpart, we extract small overlapping patches $\mathbf{P}^{HR}_i \in \mathbb{R}^{C\times p\times p}$ and $\mathbf{P}^{LR}_i \in \mathbb{R}^{C\times p' \times p'}$ from both $\mathbf{I}^{HR}$ and $\mathbf{I}^{LR}$.

Thus, for a given LR patch $\mathbf{P}^{LR}_i$, the goal is to reconstruct its corresponding HR patch $\mathbf{P}^{SR}_i \in \mathbb{R}^{p \times p \times C}$ using a learned mapping $f_{\theta}$, such that:

\begin{equation}
\mathbf{P}^{SR}_i = f_{\theta}(\mathbf{P}^{LR}_i),
\end{equation}

\noindent where $\theta$ represents the parameters of our SR model. The final SR image $\mathbf{I}^{SR}$ is obtained by aggregating the predicted patches $\mathbf{P}^{SR}_i$ across the image domain.

The patch-based strategy allows for finer local structure preservation, a better handling of complex patterns, and efficient training on smaller receptive fields. It also enables the network to focus on local dependencies and details, which are crucial for reconstructing high-frequency information from low-resolution patches. Besides, this formulation ensures that the learned model can generalize better across varying image scales.





\subsection{LPSR model overview}
\label{sec:arch}
\begin{figure}[h]
    \centering
    \includegraphics[width=\linewidth]{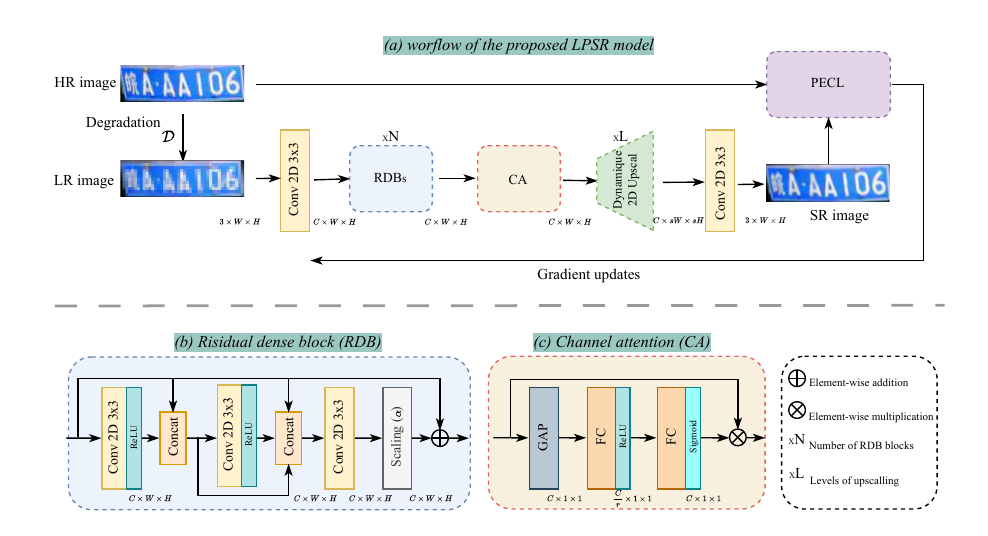}
    \caption{Workflow of the proposed LPSR model. Residual dense blocks (RDBs) capture complex hierarchical features using dense connections and residual learning. Channel attention (CA) focuses on the most informative feature channels. The pixel and embedding consistency loss (PECL) computes pixel-to-pixel and embeddings fidelity for the gradient updates.}
    \label{fig:diag_model}
\end{figure}

To achieve high-quality super-resolution, the proposed model is built upon foundational observations derived from state-of-the-art Single Image Super Resolution (SISR) techniques, specifically incorporating residual dense blocks (RDBs)~\citep{zhang2018residual} and channel attention mechanisms~\citep{chen2021attention}. The architecture is designed to progressively refine low-resolution inputs $\mathbf{P}_{\text{LR}} \in \mathbb{R}^{C\times W\times H}$ into high-resolution outputs $\mathbf{P}_{\text{SR}} \in \mathbb{R}^{C\times W\times H}$, where $C$, $W$, and $H$ denote the number of channels, width and height, respectively, through a series of strategically implemented processing stages. 

The input LR patch is first processed by a $3\times3$ convolutional layer to obtain shallow features:

\begin{equation}
\mathbf{F}_{0} = \text{Conv}_{3 \times 3}(\mathbf{P}_{\text{LR}}),
\end{equation}

\noindent where $\mathbf{F}_{0}$ represents the feature map produced by the convolution operation. The $3 \times 3$ kernel size is employed to balance local structure and computational efficiency, enabling the extraction of essential features such as edges and textures, which are critical for subsequent processing stages.

Following the initial feature extraction, the low-resolution features $\mathbf{F}_{0}$ are fed into a serial of $N$ Residual Dense Blocks (RDBs), which are designed to capture complex hierarchical features through dense connections and residual learning, see Fig.~\ref{fig:diag_model} (b). The output of the $i$-th RDB can be expressed as:

\begin{equation}
\mathbf{F}_{i} = \mathbf{F}_{i-1} + \text{RDB}(\mathbf{F}_{i-1}),
\end{equation}

\noindent where $RDB(\cdot)$ represents the operations performed within the $i$-th RDB, including convolution, activation, and feature concatenation. A crucial component of each RDB is the scaling operation applied to the output before it is added back to the residual input:

\begin{equation}
\mathbf{F}_{i} = \alpha \cdot \text{Conv}_{\text{last}}(\mathbf{F}_{i-1}) + \mathbf{F}_{i-1},
\end{equation}

\noindent where $\text{Conv}_{\text{last}}(\cdot)$ denotes the last convolution operation within $i$-th RDB and $\alpha$ is a learnable scaling factor that adjusts the contribution of the RDB output relative to the residual input. This scaling mechanism enhances the ability of the model to control the influence of each residual dense learning output, allowing for adaptive learning of the feature importance during the training process. By incorporating this scaling operation, the RDBs effectively facilitate the retention of rich information and enhance gradient flow during training, addressing challenges such as degradation and the vanishing gradient problem. 

The architecture of each RDB allows for the concatenation of feature maps from previous layers, facilitating rich information retention and enhancing the gradient flow during training~\citep{zhang2018residual}. Besides, such a design addresses the challenges associated with deep networks, such as the vanishing gradient problem, by maintaining a direct path for gradient propagation.

The Channel Attention (CA) block is pivotal in enhancing the model’s ability to prioritize the most informative feature channels~\citep{chen2021attention}, significantly boosting the overall performance of the super-resolution task. The CA mechanism operates on the feature maps generated by the final RDB block, as depicted in Fig.~\ref{fig:diag_model} (c). These feature maps are denoted as $\mathbf{F} \in \mathbb{R}^{C \times H \times W}$, where $C$ is the number of channels, and $H$ and $W$ represent the height and width of the feature maps, respectively. The CA block employs a global average pooling operation to capture the global context, resulting in a channel descriptor $z \in \mathbb{R}^{C \times 1 \times 1}$:

\begin{equation}
z = \text{GAP}(\mathbf{F}) = \frac{1}{H \times W} \sum_{i=1}^{H} \sum_{j=1}^{W} \mathbf{F}(:, i, j),
\end{equation}

\noindent where GAP denotes the global average pooling operation. This descriptor is then passed through two fully connected (FC) layers to generate the attention weights. The first layer reduces the dimensionality of the channel descriptor:

\begin{equation}
z_{1} = \text{ReLU}(\theta_{\text{FC1}} \cdot z + b_{1}),
\end{equation}

\noindent where $\theta_{\text{FC1}} \in \mathbb{R}^{\frac{C}{r} \times C}$ is the weight matrix of the first FC layer, $b_{1}$ is the bias term, and $r$ is the reduction ratio. The output of this layer is then passed through the second fully connected layer to restore the original dimensionality:

\begin{equation}
z_{2} = \sigma(\theta_{\text{FC2}} \cdot z_{1} + b_{2}),
\end{equation}

\noindent where $\theta_{\text{FC2}} \in \mathbb{R}^{C \times \frac{C}{r}}$ and $b_{2}$ are the weight matrix and bias of the second FC layer, respectively, and $\sigma$ represent the sigmoid activation function. The resulting attention vector $z_{2} \in \mathbb{R}^{C}$ is then reshaped and used to scale the original feature maps:

\begin{equation}
\mathbf{F}_{\text{CA}} = \mathbf{F} \otimes z_{2},
\end{equation}

\noindent where $\mathbf{F}_{\text{CA}}$ represents the output of the CA block after applying the attention weights. This block enhances the network’s representational capacity by emphasizing informative channels and suppressing less relevant ones, resulting in improved feature extraction for subsequent processing stages. Integrating the CA block into the super-resolution architecture is crucial, as it aligns with the model’s goal of generating high-quality images by selectively focusing on critical features that significantly impact perceptual quality.

Subsequently, the proposed model incorporates a dynamic upsampling mechanism through a series of 2D transposed convolution operations~\citep{8618415}. This approach incrementally doubles the spatial dimensions of the feature maps, effectively enhancing the spatial resolution while preserving essential details. The multi-stage upsampling strategy enables the model to progressively refine and upscale the output patches across multiple levels, which is critical for mitigating artifacts and improving overall image fidelity. Therefore, the upscaled features $\mathbf{F}_{\text{UP}} \in \mathbb{R}^{C\times sW\times sH}$ are obtained by:

\begin{equation}
    \mathbf{F}_{\text{UP}} = \text{ReLU} \left( \text{ConvT}^L(\mathbf{F}_{\text{CA}}) \right),
\end{equation}

\noindent where $\mathbf{F}_{\text{CA}}$ represents the feature maps obtained from the channel attention block. $L$ denotes the number of upsampling stages required, calculated as $L=\log_2(s)$, with $s$ being the scale factor. The use of transposed convolutions $\text{ConvT}$ allows for the integration of learned features from the preceding layers, ensuring that the generated high-resolution output retains the rich structural and contextual information from the low-resolution input.

Finally, the model ends with a convolutional layer that further refines the output feature maps to produce the final SR patch. This layer utilizes a $3 \times 3$ convolution operation to seamlessly integrate the features extracted in the preceding stages into a coherent and high-fidelity output. The final SR patch is obtained as:

\begin{equation}
    \mathbf{P}_{\text{SR}} = \text{Conv}(\mathbf{F}_{\text{up}}),
\end{equation}

\noindent where $\mathbf{P}_{\text{SR}}$ denotes the super-resolved patch, and $\text{Conv}$ represents the $3 \times 3$ convolution operation applied to the upsampled feature maps $\mathbf{F}_{\text{up}}$. 

This final convolution serves multiple purposes. It consolidates the features learned through the RDBs and the CA mechanism, effectively integrating high-level representations with spatial information. Additionally, it plays a crucial role in mitigating potential artifacts introduced during the upsampling stages, ensuring that the output not only achieves the target resolution but also maintains visual consistency and high quality.

\subsection{Pixel and embedding consistency loss}

\begin{figure}[h]
    \centering
    \includegraphics[width=\linewidth]{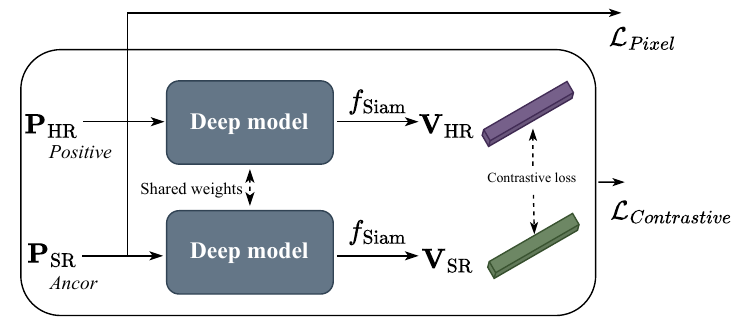}
    \caption{Illustration of the Siamese network architecture used for pixel and embedding consistency loss. The network comprises two identical sub-networks extracting embeddings from the super-resolved ($\mathbf{P}_{\text{SR}}$) and high-resolution ($\mathbf{P}_{\text{HR}}$) patches. The distance $D$ between the embeddings $\mathbf{V}_{\text{SR}}$ and $\mathbf{V}_{\text{HR}}$ is computed in the embedding space, informing the Contrastive loss $\mathcal{L}_{\text{Contrastive}} = \max(m - D, 0)^2$, which reinforces the model's capacity to preserve key features across resolutions.}
    \label{fig:diag_loss}
\end{figure}

To effectively enhance the performance of the LPSR model, we propose a comprehensive loss function that combines pixel-wise loss with embedding similarity learning through a Siamese network architecture~\citep{koch2015siamese}. The pixel and embedding consistency loss (PECL) function $\mathcal{L}_{\text{PEC}}$ is formulated as a weighted sum of the mean squared error (MSE) loss $\mathcal{L}_{\text{pixel}}$ and the Contrastive loss $\mathcal{L}_{\text{contrastive}}$, which incorporates the embedding similarity component.

\textbf{Pixel-wise loss}: it quantifies the discrepancy between the super-resolved output $\mathbf{P}_{\text{SR}}$ and the corresponding high-resolution target $\mathbf{P}_{\text{HR}}$. This loss is computed using the MSE, expressed as:

\begin{equation}
\mathcal{L}_{\text{pixel}} = \frac{1}{K} \sum_{i=1}^{K} (\mathbf{P}_{\text{SR}}^{(i)} - \mathbf{P}_{\text{HR}}^{(i)})^2,
\end{equation}

\noindent where $K$ represents the total number of pixels in the image patch, and $\mathbf{P}_{\text{SR}}^{(i)}$ and $\mathbf{P}_{\text{HR}}^{(i)}$ denote the pixel values at the $i$-th position for the super-resolved and high-resolution patches, respectively. This loss function effectively captures the average squared differences between corresponding pixel values, promoting fine-grained accuracy in pixel representation.

\textbf{Contrastive loss}: To ensure that the super-resolved patches maintain key features characteristic of the high-resolution patches, we implement a Siamese network~\citep{koch2015siamese} to extract embeddings from both $\mathbf{P}_{\text{SR}}$ and $\mathbf{P}_{\text{HR}}$. A Siamese network comprises two identical sub-networks that share the same architecture and parameters, allowing for direct comparison of the generated embeddings. 

The Siamese network architecture leverages a pre-trained ResNet-18 model, with the final fully connected layer replaced to yield embeddings of size 128. For an input pair $(\mathbf{P}_{\text{SR}}, \mathbf{P}_{\text{HR}})$ (as shown in Fig.~\ref{fig:diag_loss}), the network outputs two embeddings:

\begin{equation}
\mathbf{V}_{\text{SR}}, \mathbf{V}_{\text{HR}} = f_{\text{Siam}}(\mathbf{P}_{\text{SR}}, \mathbf{P}_{\text{HR}}; \theta_{\text{Siam}}, d),
\end{equation}

\noindent where $\theta_{\text{Siam}}$ represents the parameters of the Siamese network and $d$ the size of the embedding. To maintain consistent distance magnitudes across samples and prevent the embeddings from growing arbitrarily large, an L2 normalization step is applied to the output embedding. This normalization constrains the embeddings to a unit hypersphere, which improves convergence and training stability. Therefore, the normalized embedding $\tilde{\mathbf{V}}$ is given by:
\begin{equation}
\tilde{\mathbf{V}} = \frac{\mathbf{V}}{\|\mathbf{V}\|_2},
\end{equation}

\noindent where $\|\mathbf{V}\|_2$ is the L2 norm of $\mathbf{V}$. This ensures that all embeddings lie on a consistent scale.

The objective of this architecture is to generate similar embeddings for the super-resolved and high-resolution patches, reflecting their inherent similarity. To achieve this, we employ the Contrastive loss function $\mathcal{L}_{\text{Contrastive}}$, which encourages similarity in the embedding space. This loss is simplified to:

\begin{equation}
\mathcal{L}_{\text{Contrastive}} = \max(m - D, 0)^2,
\end{equation}

\noindent where $D$ is the Manhattan distance ($\ell_1$-norm) between the embeddings of the super-resolved and high-resolution patches:

\begin{equation}
D = \| \tilde{\mathbf{V}}_{\text{SR}} -  \tilde{\mathbf{V}}_{\text{HR}} \|_1.
\end{equation}

\noindent In this context:
\begin{itemize}
    \item $m$ represents a margin, a hyperparameter that establishes a threshold distance between the embeddings, set to 2 in this study.

    \item The squaring operation $(m - D)^2$ imposes a heavier penalty for larger deviations, thereby reinforcing the model's capacity to minimize the distance $D$ when it is below the margin $m$.
\end{itemize}

The Manhattan distance ($\ell_1$-norm) offers several advantages for measuring embedding similarity, particularly in high-dimensional feature spaces. Unlike the Euclidean distance for instance, which squares differences and can amplify the influence of outliers, the Manhattan distance computes the sum of absolute differences, making it more robust to noisy or extreme feature values. This property is beneficial when embeddings exhibit sparsity or when certain dimensions dominate due to variability in the data. Additionally, the Manhattan distance treats each feature dimension independently, which aligns well with many neural embedding spaces where feature contributions vary. Providing stable gradients also facilitates smoother optimization during training, improving alignment and generalization~\citep{aggarwal2001surprising, gohrani2019different}.

The focus on embedding similarity loss is particularly relevant for applications such as optical character recognition (OCR)~\citep{Chaudhuri2017}, automatic number plate recognition (ANPR)~\citep{8748287}, and vehicle identification. By forcing the model to minimize the distance between embeddings of super-resolved and high-resolution images, the embedding similarity loss ensures that the reconstructed images align more closely with their high-resolution counterparts in the feature space. This alignment is crucial for preserving distinctive features necessary for accurate recognition and identification, as it guarantees that critical details are reconstructed, enhancing fidelity and perceptual quality.

\textbf{Total loss}: The total loss function, $\mathcal{L}_{\text{PECL}}$ is designed as a weighted sum of two complementary components: the pixel-wise loss ($L_{\text{pixel}}$) and the contrastive loss ($L_{\text{contrastive}}$). The pixel-wise loss ensures fidelity at the pixel level by minimizing the MSE. In contrast, the contrastive loss focuses on aligning the embeddings of the predicted super-resolved image and the high-resolution target in the feature space, minimizing the distance between their representations. Together, these losses enforce both pixel-level accuracy and feature-level consistency, improving the perceptual and semantic quality of the reconstructed images. The total loss is expressed as:

\begin{equation}
\mathcal{L}_{\text{PECL}} = w_{\text{pixel}} \cdot \mathcal{L}_{\text{pixel}} + w_{\text{contrastive}} \cdot \mathcal{L}_{\text{contrastive}},
\end{equation}

\noindent where $w_{\text{pixel}}$ and $w_{\text{contrastive}}$ are the weights that govern the contributions of the pixel-wise and contrastive losses, respectively. These weights are learnable parameters constrained within the range $0, 1$ to ensure balanced optimization. Furthermore, they are designed to satisfy $w_{\text{pixel}} + w_{\text{contrastive}} = 1$, ensuring that the total weight is dynamically distributed between the two loss components. To maintain valid weight ranges, $\mathcal{L}_{\text{PECL}}$ clips $w_{\text{pixel}}$ and $w_{\text{contrastive}}$ within $0, 1$ during training. This ensures stability and prevents either loss component from dominating excessively. By enforcing this balance, the total loss function effectively combines the strengths of pixel-level fidelity and feature-level alignment, enabling the model to achieve sharper reconstructions while preserving semantic consistency.




\section{Experiments}

\subsection{Datasets and Implementation Details}

\textbf{Dataset:} \textcolor{black}{To assess the performance of our proposed approach across diverse real-world scenarios, we use two publicly available datasets: the Chinese City Parking Dataset (CCPD)\citep{xu2018towards} and the PKU Vehicle Dataset\citep{7752971}. 
\begin{description}
    \item[CCPD~\citep{xu2018towards}:] This dataset comprises over 200k images of license plates captured under varying real-world conditions, including different angles, distances, and lighting scenarios. Its diversity makes it particularly well-suited for our study, as it presents significant challenges for traditional recognition systems. We use 100,000 images for training and 1k for testing and validation, ensuring well-balanced sets for both learning and evaluation.
    \item[PKU~\citep{7752971}:] This dataset consists of over 4k license plate images taken in diverse environments, such as highways under normal daylight and nightlight conditions and intersections with crosswalks during both day and night. We randomly select 3,5k images for training and allocate the remaining for testing, ensuring representative data distributions.
\end{description}
}

\begin{figure}[h]
    \centering
    \includegraphics[width=\linewidth]{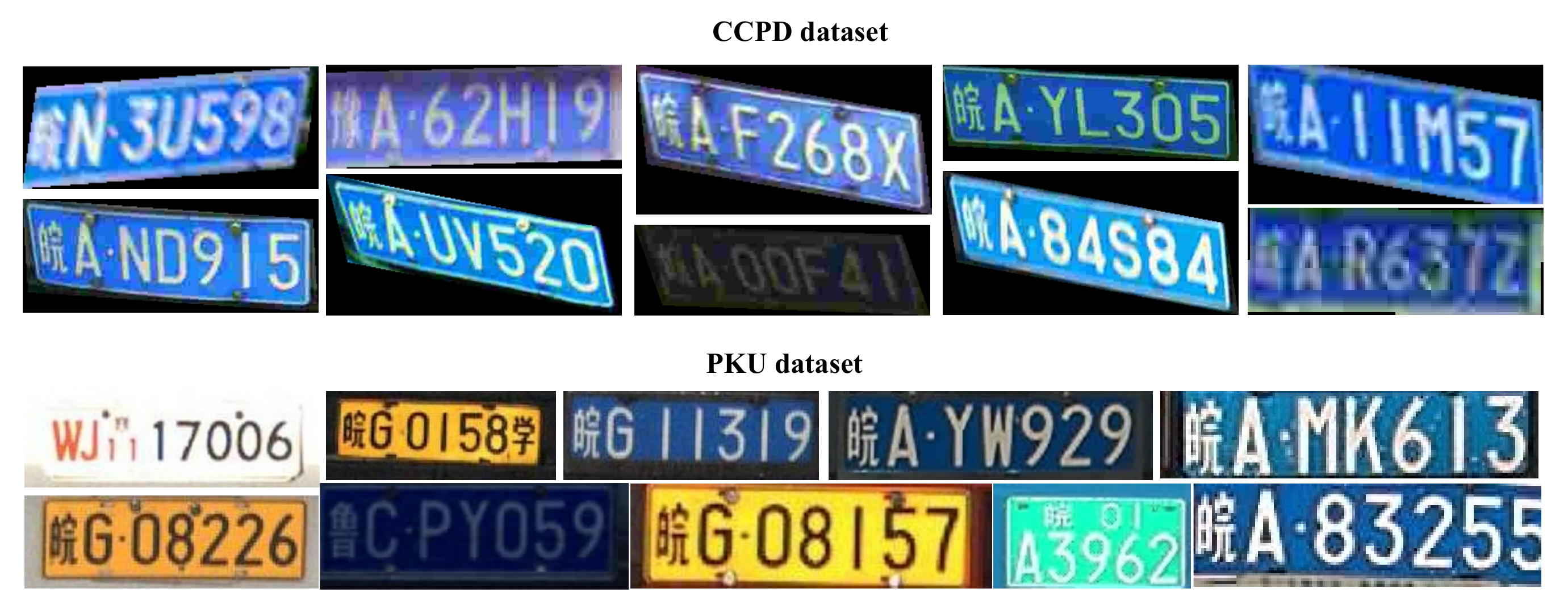}
    \caption{Examples from the training dataset showcasing diverse capturing conditions, including variations in angles, distances, and lighting.}
    \label{fig:training_exp}
\end{figure}

The training datasets exhibit diverse capturing conditions, as illustrated in Fig.~\ref{fig:training_exp}, including variations in angles, distances, and lighting. Additionally, many of the HR images suffer from significant distortions, further complicating the super-resolution task and emphasizing the need for robust models capable of handling degraded inputs.

\begin{figure}[h]
    \centering
    \includegraphics[width=\linewidth]{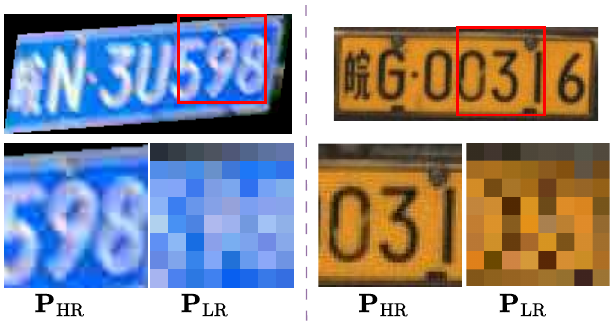}
    \caption{\textcolor{black}{Examples of training patches and the applied degradation process, highlighting the downscaling operation used to simulate low-resolution inputs. (left) example from CCPD and (right) from PKU.}}
    \label{fig:exp_patches}
\end{figure}

As described in Sec.\ref{sec:arch}, the LPSR model processes input patches instead of full images. This patch-based design is a widely adopted practice in super-resolution (SR) tasks\citep{dong2015image, liang2021swinir, wang2021real}, as it enhances computational efficiency while enabling the model to focus on localized features. Moreover, it effectively handles the inherent variability in resolution and aspect ratios of real-world license plate images, ensuring robust generalization across diverse input conditions. Fig.~\ref{fig:exp_patches} showcases sample pairs of high-resolution (HR) and low-resolution (LR) patches. Given that the original images already exhibit visual degradations, the degradation process $\mathcal{D}$ is limited to downscaling. In this work, we address the challenging task of $\times 8$ scaling, categorized as extreme super-resolution due to the significant loss of visual information in LR inputs. \textcolor{black}{For this study, the patch size is set $64 \times 64$ pixels.}

\vspace{1cm}
\textbf{Implementation Details:} The LPSR model is implemented using the PyTorch library \citep{pytorch} and trained on a server equipped with an Intel Xeon Silver 4208 2.1GHz CPU, 192GB of RAM, and an Nvidia Tesla V100S GPU with 32GB of memory. We train the model for 1000k iterations with a batch size of 128. We employ the Adam optimizer~\citep{kingma2014adam} to update the parameters of the model, with an initial learning rate $1e-4$. To adaptively adjust the learning rate throughout the training process, we utilize a cosine annealing learning rate scheduler~\citep{loshchilov2017sgdr}. The scheduler gradually decreases the learning rate over the iterations following a cosine curve. This dynamic adjustment aids in fine-tuning the model's parameters as training progresses.


\label{sec:cret}
\textcolor{black}{
\textbf{Evaluation criteria:} The evaluation of LPSR models focuses on two key aspects: visual quality and character recognition performance. Visual quality assesses how closely the SR images resemble the ground truth HR images, ensuring the restored images are perceptually faithful to the originals. \textcolor{black}{For this purpose, we use peak signal-to-noise ratio (PSNR), structural similarity index (SSIM)~\citep{wang2004image}, and learned perceptual image patch similarity (LPIPS)~\citep{zhang2018unreasonable}. PSNR and SSIM are measured on both full-color RGB channels (denoted as $PSNR$ and $SSIM$) and luminance channel (denoted as $PSNR_Y$ and $SSIM_Y$), with the latter being particularly significant due to its strong correlation with human visual perception.} In addition to visual fidelity, the effectiveness of the super-resolution model in enhancing character readability is evaluated using an OCR. PaddleOCR~\citep{paddleocr}, a widely-used open-source OCR system~\citep{peng2024paddle, sarkar2024automatic, ma2024application, kulkarni2023helmet}, is employed to read characters on the SR images. The performance is quantified using several metrics, including 1) exact match accuracy (EMA) to quantify the percentage of cases where the predicted LP text exactly matches the ground truth, 2) The Average Levenshtein similarity (L-sim)~\citep{levenshtein1966binary}, which provides a measure of partial correctness by computing the edit distance between predictions and references, and 3) character error rate (CER) and word error rate (WER) assess the frequency of misrecognized characters and words, respectively. In addition, to capture performance at a more granular level, average character-level precision (Prec.), recall, and F1-scores (F1-S) are computed, reflecting the accuracy and consistency of individual character predictions. As these metrics depend on ground truth annotations, the OCR-based evaluation is conducted exclusively on the CCPD dataset as it provides such annotations.}

\subsection{Results and discussion}

\subsubsection{Performance comparison}
\label{sec:perf_comp}

The performance of the proposed SR model is evaluated in comparison to the baseline Bicubic and SOTA methods, including SRCNN~\citep{dong2015image}, MSRN~\citep{li2018multi}, ESPCN~\citep{shi2016real}, ESRGAN~\citep{wang2018esrgan}, TBSRN~\citep{chen2021scene}, and SwinIR~\citep{liang2021swinir}. Table \ref{table:performance_comparison} presents the results of this comparison, where the median values, along with the standard deviations, are reported over the testing set. \textcolor{black}{In addition to quality scores,  Table \ref{table:performance_comparison} reports the complexity analysis quantified by means of number of parameters (\#P) and floating point operations per second (GFLOPs).}


\begin{table}[htbp]
\centering
\caption{Performance comparison of the proposed model with various SR methods on the CCPD dataset~\citep{xu2018towards}. The median (\textcolor{gray}{$\pm$ standard deviation}) over the testing set is reported. The best and second-best performances are respectively highlighted in \textcolor{red}{\textbf{bold red}} and \textcolor{blue}{\textbf{bold blue}}.}
\rowcolors{0}{gray!10}{white} 
\begin{tabular}{@{}l|p{2.6cm}|p{2.8cm}|p{1.5cm}|p{1.5cm}@{}}
\toprule
\textbf{SR Method}           & \textbf{PSNR/PSNRy ($\uparrow$)} & \textbf{SSIM/SSIMy ($\uparrow$)} & \textbf{LPIPS ($\downarrow$)} & \textbf{\#P / GFLOPs} \\ \midrule
\textbf{Bicubic}             & 18.53 (\textcolor{gray}{$\pm$ 2.78}) / 18.70 (\textcolor{gray}{$\pm$ 2.82})  & 0.4504 (\textcolor{gray}{$\pm$ 0.11}) / 0.4619 (\textcolor{gray}{$\pm$ 0.12}) & 0.395 (\textcolor{gray}{$\pm$ 0.07}) & - \\ \midrule
\textbf{SRCNN~\citep{dong2015image}}  & 19.57 (\textcolor{gray}{$\pm$ 2.72}) / 19.84 (\textcolor{gray}{$\pm$ 2.78}) & 0.5426 (\textcolor{gray}{$\pm$ 0.10}) / 0.5675 (\textcolor{gray}{$\pm$ 0.11}) & 0.319 (\textcolor{gray}{$\pm$ 0.11}) & 57K / 0.28 \\ \midrule
\textbf{MSRN~\citep{li2018multi}}    & 19.78 (\textcolor{gray}{$\pm$ 2.74}) / 19.84 (\textcolor{gray}{$\pm$ 2.82}) & 0.5069 (\textcolor{gray}{$\pm$ 0.10}) / 0.5237 (\textcolor{gray}{$\pm$ 0.11}) & 0.362 (\textcolor{gray}{$\pm$ 0.07})& 6M / 15.11 \\ \midrule
\textbf{ESPCN~\citep{shi2016real}}  & 23.52 (\textcolor{gray}{$\pm$ 2.47}) /  24.11 (\textcolor{gray}{$\pm$ 0.61}) & 0.7383 (\textcolor{gray}{$\pm$ 0.06}) / 0.7750 (\textcolor{gray}{$\pm$ 0.07}) & 0.146 (\textcolor{gray}{$\pm$ 0.04})  & 800K / 14.50 \\ \midrule
\textbf{ESRGAN~\citep{wang2018esrgan}}  & 18.76 (\textcolor{gray}{$\pm$ 2.26}) / 19.01 (\textcolor{gray}{$\pm$ 2.33}) & 0.5709 (\textcolor{gray}{$\pm$ 0.08}) / 0.5990 (\textcolor{gray}{$\pm$ 0.08}) & \textcolor{blue}{\textbf{0.122}} (\textcolor{gray}{$\pm$ 0.04}) & 16M / 17.64 \\ \midrule
\textbf{TBSRN~\citep{chen2021scene}}  & \textcolor{blue}{\textbf{23.76}} (\textcolor{gray}{$\pm$ 2.48}) / \textcolor{blue}{\textbf{24.42}} (\textcolor{gray}{$\pm$ 2.63}) & \textcolor{blue}{\textbf{0.7480}} (\textcolor{gray}{$\pm$ 0.07}) / \textcolor{blue}{\textbf{0.7854}} (\textcolor{gray}{$\pm$ 0.08}) & 0.143 (\textcolor{gray}{$\pm$ 0.05})  & 12M / 18.49 \\ \midrule
\textbf{SwinIR~\citep{liang2021swinir}} & 23.56 (\textcolor{gray}{$\pm$ 2.50}) / 24.18 (\textcolor{gray}{$\pm$ 2.65}) & 0.7477 (\textcolor{gray}{$\pm$ 0.06}) / 0.7816 (\textcolor{gray}{$\pm$ 0.07}) & 0.147 (\textcolor{gray}{$\pm$ 0.05})  & 11M / 46.45 \\ \midrule
\textbf{Ours}              & \textcolor{red}{\textbf{25.13}} (\textcolor{gray}{$\pm$ 2.46})  / \textcolor{red}{\textbf{25.92}} (\textcolor{gray}{$\pm$ 2.62})  & \textcolor{red}{\textbf{0.8127}} (\textcolor{gray}{$\pm$ 0.06}) / \textcolor{red}{\textbf{0.8458}} (\textcolor{gray}{$\pm$ 0.06}) & \textcolor{red}{\textbf{0.106}} (\textcolor{gray}{$\pm$ 0.04})  & 1.9M / 13.35 \\ \bottomrule
\end{tabular}\label{table:performance_comparison}

\end{table}

\begin{table}[htbp]
\centering
\caption{Performance comparison of the proposed model with various SR methods on the PKU dataset~\citep{7752971}. The median (\textcolor{gray}{$\pm$ standard deviation}) over the testing set is reported. The best and second-best performances are respectively highlighted in \textcolor{red}{\textbf{bold red}} and \textcolor{blue}{\textbf{bold blue}}.}
\rowcolors{0}{gray!10}{white} 
\begin{tabular}{@{}l|p{2.6cm}|p{2.8cm}|p{1.5cm}|p{1.5cm}@{}}
\toprule
\textbf{SR Method}           & \textbf{PSNR/PSNRy ($\uparrow$)} & \textbf{SSIM/SSIMy ($\uparrow$)} & \textbf{LPIPS ($\downarrow$)} & \textbf{\#P / GFLOPs} \\ \midrule
\textbf{Bicubic}             & 15.89 (\textcolor{gray}{$\pm$ 1.95}) /  15.98 (\textcolor{gray}{$\pm$ 1.98})  & 0.2832 (\textcolor{gray}{$\pm$ 0.09}) / 0.2839 (\textcolor{gray}{$\pm$ 0.09}) &  0.578 (\textcolor{gray}{$\pm$ 0.07}) & - \\ \midrule
\textbf{SRCNN~\citep{dong2015image}}  & 15.37 (\textcolor{gray}{$\pm$ 1.95}) / 15.41 (\textcolor{gray}{$\pm$ 1.92}) &  0.3267 (\textcolor{gray}{$\pm$ 0.09}) / 0.3341 (\textcolor{gray}{$\pm$ 0.10}) & 0.465 (\textcolor{gray}{$\pm$ 0.08}) & 57K / 0.28 \\ \midrule
\textbf{MSRN~\citep{li2018multi}}    & 15.58 (\textcolor{gray}{$\pm$ 2.17}) / 15.61 (\textcolor{gray}{$\pm$ 2.15}) & 0.3132 (\textcolor{gray}{$\pm$ 0.10}) / 0.3185 (\textcolor{gray}{$\pm$ 0.11}) & 0.4442 (\textcolor{gray}{$\pm$ 0.09})& 6M / 15.11 \\ \midrule
\textbf{ESPCN~\citep{shi2016real}}  & 17.52 (\textcolor{gray}{$\pm$ 2.99}) /  17.61 (\textcolor{gray}{$\pm$ 2.15}) & 0.5492 (\textcolor{gray}{$\pm$ 0.12}) / 0.5743 (\textcolor{gray}{$\pm$ 0.12}) & 0.199 (\textcolor{gray}{$\pm$ 0.06})  & 800K / 14.50 \\ \midrule
\textbf{ESRGAN~\citep{wang2018esrgan}}  & \textcolor{blue}{\textbf{18.74}} (\textcolor{gray}{$\pm$ 3.59}) / \textcolor{blue}{\textbf{18.92}} (\textcolor{gray}{$\pm$ 3.69}) & \textcolor{blue}{\textbf{0.6796}} (\textcolor{gray}{$\pm$ 0.13}) / \textcolor{blue}{\textbf{0.7014}} (\textcolor{gray}{$\pm$ 0.13}) & \textcolor{blue}{\textbf{0.110}} (\textcolor{gray}{$\pm$ 0.05}) & 16M / 17.64 \\ \midrule
\textbf{TBSRN~\citep{chen2021scene}}  & 17.31 (\textcolor{gray}{$\pm$ 3.17}) / 17.47 (\textcolor{gray}{$\pm$ 3.22}) & 0.5578 (\textcolor{gray}{$\pm$ 0.13}) / 0.5813 (\textcolor{gray}{$\pm$ 0.13}) & 0.174 (\textcolor{gray}{$\pm$ 0.06})  & 12M / 18.49 \\ \midrule
\textbf{SwinIR~\citep{liang2021swinir}} & 17.39 (\textcolor{gray}{$\pm$ 3.26}) / 17.62 (\textcolor{gray}{$\pm$ 3.31}) & 0.5414 (\textcolor{gray}{$\pm$ 0.14}) / 0.5672 (\textcolor{gray}{$\pm$ 0.15}) & 0.188 (\textcolor{gray}{$\pm$ 0.06})  & 11M / 46.45 \\ \midrule
\textbf{Ours}              & \textcolor{red}{\textbf{19.26}} ($\pm$\textcolor{gray}{3.35}) / \textcolor{red}{\textbf{19.46}} ($\pm$\textcolor{gray}{3.77}) & \textcolor{red}{\textbf{0.7009}} ($\pm$\textcolor{gray}{0.12}) / \textcolor{red}{\textbf{0.7222}} ($\pm$\textcolor{gray}{0.12}) & \textcolor{red}{\textbf{0.101}} ($\pm$\textcolor{gray}{0.04})  & 1.9M / 13.35 \\ \bottomrule
\end{tabular}\label{table:performance_comparison}

\end{table}

\begin{figure}[htbp]
    \centering
    \includegraphics[width=\linewidth]{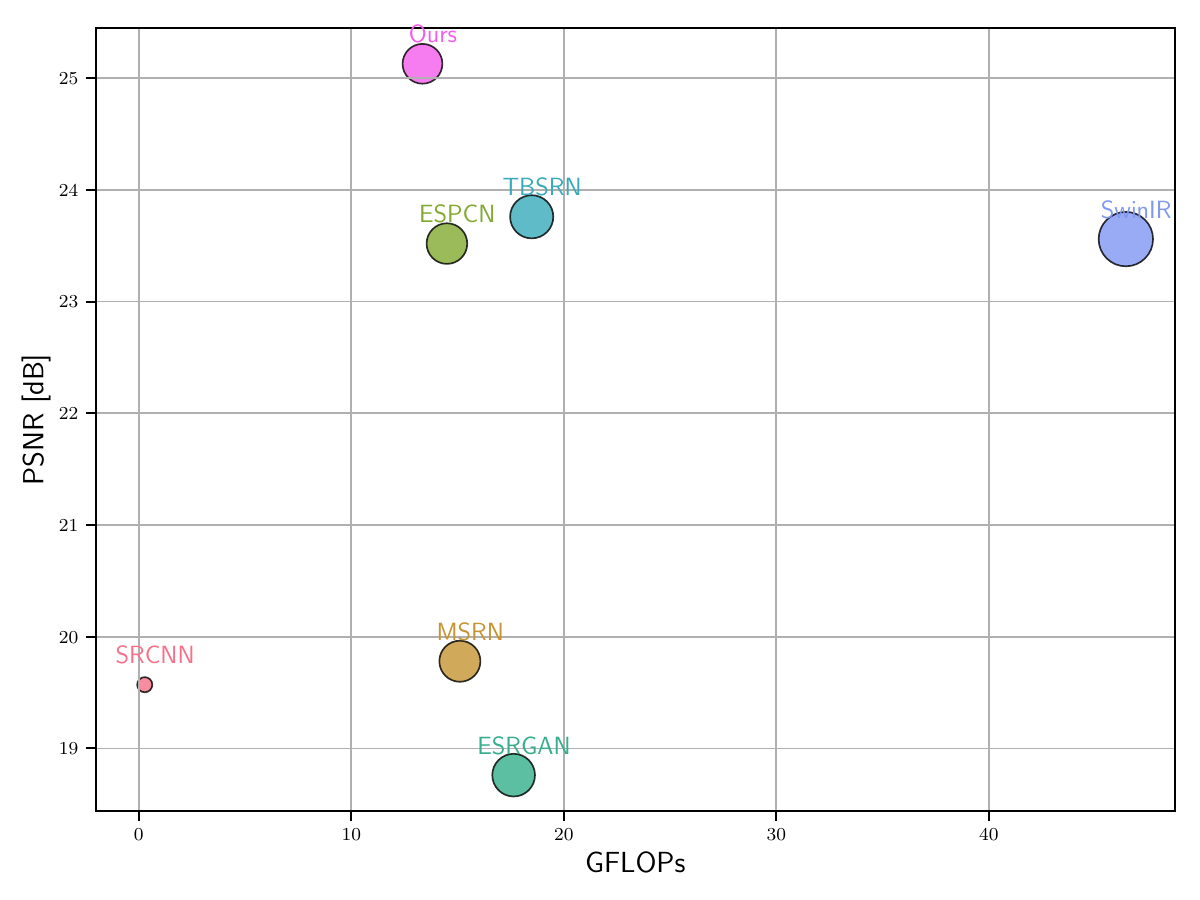}
    \caption{\textcolor{black}{GFLOPs vs. PSNR on the CCPD dataset~\citep{xu2018towards} for the proposed method compared to SOTA models. The scatter plot visualizes the trade-off between computational complexity (measured in GFLOPs) and reconstruction quality (measured in PSNR) for different SR models. Each point represents a model, with a radius proportional to the square root of its GFLOPs. The GFLOPs were computed using $64\times64$ input images and averaged over 10 iterations for consistency. }}
    \label{fig:psnr_gflops}
\end{figure}

\textbf{Quantitative comparison:}
\textcolor{black}{
On the CCPD dataset, the performance trend clearly favors advanced deep learning methods over traditional interpolation and early CNN-based approaches. Bicubic interpolation, along with SRCNN and MSRN, delivers modest PSNR values in the range of 18–20 dB and relatively low SSIM scores, reflecting their limited capacity to reconstruct fine details and textures from low-resolution inputs. In contrast, methods such as ESPCN, ESRGAN, TBSRN, and SwinIR exhibit marked improvements, with ESPCN achieving a PSNR of 23.52 dB and TBSRN and SwinIR reaching similar high-quality scores. Notably, the proposed model surpasses all competitors by achieving a PSNR of 25.13 dB (with a PSNRy of 25.92 dB) and an SSIM of 0.8127/0.8458, along with the lowest LPIPS of 0.106. These superior scores, combined with a lightweight architecture (1.9M parameters and 13.35 GFLOPs), underscore the model’s ability to enhance perceptual quality, preserve structural details, and effectively handle the diverse imaging conditions inherent in the large-scale CCPD dataset.}

\textcolor{black}{
In comparison, the PKU dataset, characterized by its smaller scale and generally lower inherent image quality, presents a more challenging testbed for super-resolution methods. On this dataset, methods such as Bicubic, SRCNN, and MSRN yield even lower PSNR values (around 15–16 dB) and SSIM scores, while modern deep learning techniques show improved performance. For instance, ESPCN reaches a PSNR of approximately 17.5 dB and an SSIM close to 0.55, and ESRGAN, highlighted as the second-best performance, attains a PSNR of 18.74 dB and an SSIM of 0.6796. However, the proposed model again takes the lead with a PSNR of 19.26 dB (PSNRy of 19.46 dB), an SSIM of 0.7009/0.7222, and the lowest LPIPS of 0.101. This consistency in outperforming state-of-the-art methods on both datasets demonstrates the robustness and adaptability of the proposed approach across varying scales and quality levels, effectively balancing high-fidelity reconstruction with computational efficiency.}

\begin{figure}[h]
    \centering
    \setlength{\tabcolsep}{0pt}
    \begin{tabular}{cc}
        \includegraphics[width=0.5\linewidth]{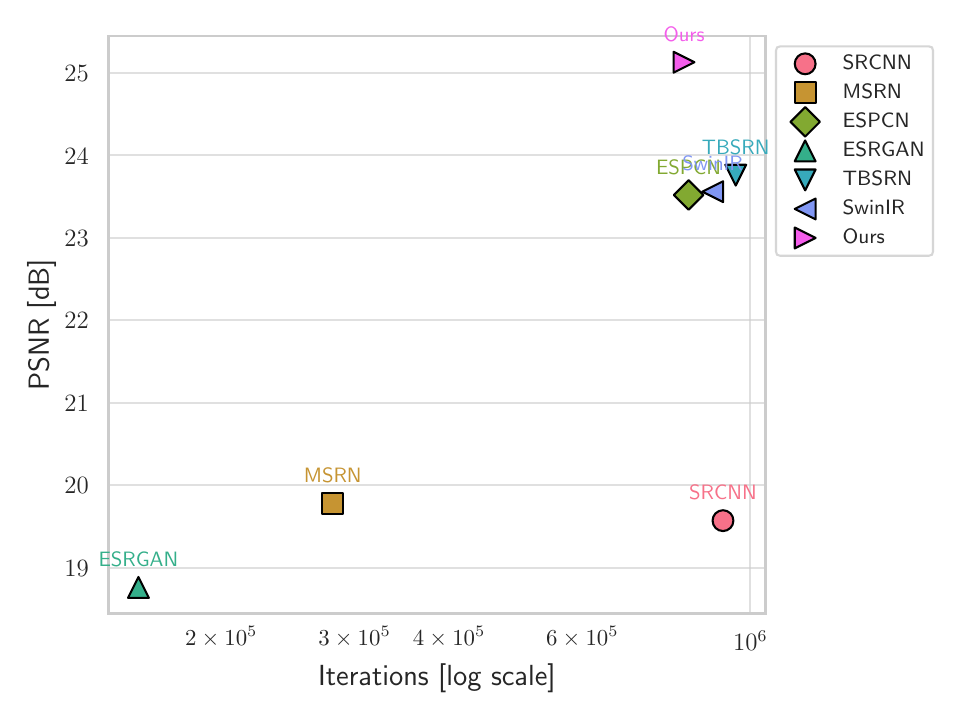} & 
        \includegraphics[width=0.5\linewidth]{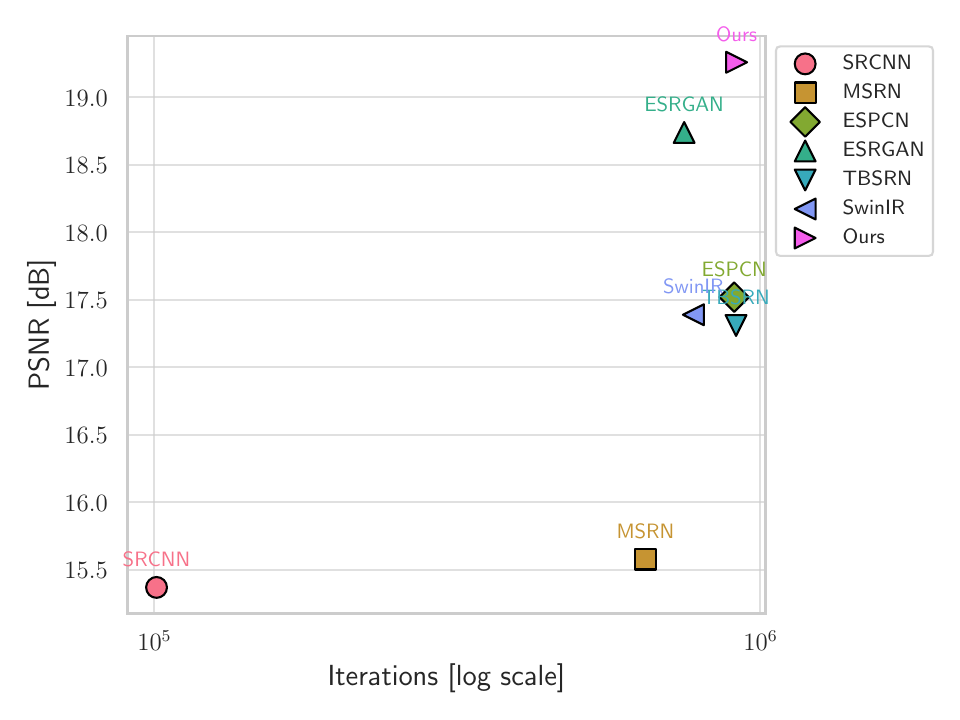}\\
    \end{tabular}
    \caption{Performance in terms of PSNR \textit{vs.} the number of iterations (on a logarithmic scale) required for convergence during training. (left) CCPD dataset~\citep{xu2018towards} and (right) PKU dataset~\citep{7752971}.}
    \label{fig:psnr_iter}
\end{figure}

\textcolor{black}{
Additionally, the analysis of computational complexity in terms of GFLOPs versus PSNR, as depicted in Fig.~\ref{fig:psnr_gflops}, reveals that higher computational complexity does not always guarantee better performance. While models like SwinIR (46.45 GFLOPs, 23.56 PSNR) exhibit high computational demands, they do not achieve the best PSNR. In contrast, TBSRN (18.49 GFLOPs, 23.76 PSNR) and ESPCN (14.5 GFLOPs, 23.52 PSNR) reach comparable if not better performance with significantly lower GFLOPs, highlighting the efficiency of certain architectures. Additionally, the trend suggests diminishing returns beyond approximately 15-20 GFLOPs, where increased computation does not necessarily yield substantial PSNR improvements. For example, ESRGAN (17.64 GFLOPs, 18.76 PSNR) has high computational cost but relatively low PSNR, likely due to its focus on perceptual quality rather than pixel-wise accuracy. Notably, our model (13.35 GFLOPs, 25.13 PSNR) achieves the highest PSNR while maintaining moderate computational complexity, indicating an efficient design. This suggests that optimized network architectures and better feature representations can significantly improve performance without excessive computational cost.} 

We further analyzed the performances in terms of PSNR on the testing set versus the number of iterations required for each model to converge. All models were trained for 1000k iterations, with the best version saved at the highest PSNR during training. The plot in Fig.~\ref{fig:psnr_iter} illustrates the PSNR \textit{vs.} number of iterations (on a logarithmic scale) on both dataset. A clear correlation between the number of iterations and the resulting PSNR values can be observed, except for SRCNN. 

Models that required more iterations to converge generally delivered higher PSNR on the testing set. This trend highlights the importance of longer training for deeper models. For instance, ESPCN, TBSRN, and SwinIR required approximately 900k iterations to converge, achieving PSNR values of 23.52 dB and 23.76 dB, respectively. These models' performances indicate their ability to capture complex features from the dataset directly correlates with their extended training duration. Similar behavior can be seen for the proposed model. MSRN and ESRGAN converged more quickly with fewer iterations (approximately 281k and 155k iterations, respectively), and achieved lower PSNR values (19.78 dB and 18.76 dB, respectively). This suggests that despite their faster convergence, these models struggled to learn the necessary low-level features for the LPSR task from the training set, which resulted in a suboptimal generalization on the testing set. Regarding SRCNN behavior, it simple architecture, which lacks the capacity for learning complex features for extreme SR, limits its performance, even with prolonged training

These observations emphasize the trade-off between convergence speed and model performance in image super-resolution tasks. Models that require longer training tend to produce better results on unseen data, confirming the need for adequate training to fully exploit the potential of more complex architectures. In contrast, quicker-converging models may face challenges in terms of learning the finer details, potentially leading to a degradation in performance.

\begin{figure}[h]
    \centering
    \includegraphics[width=\linewidth]{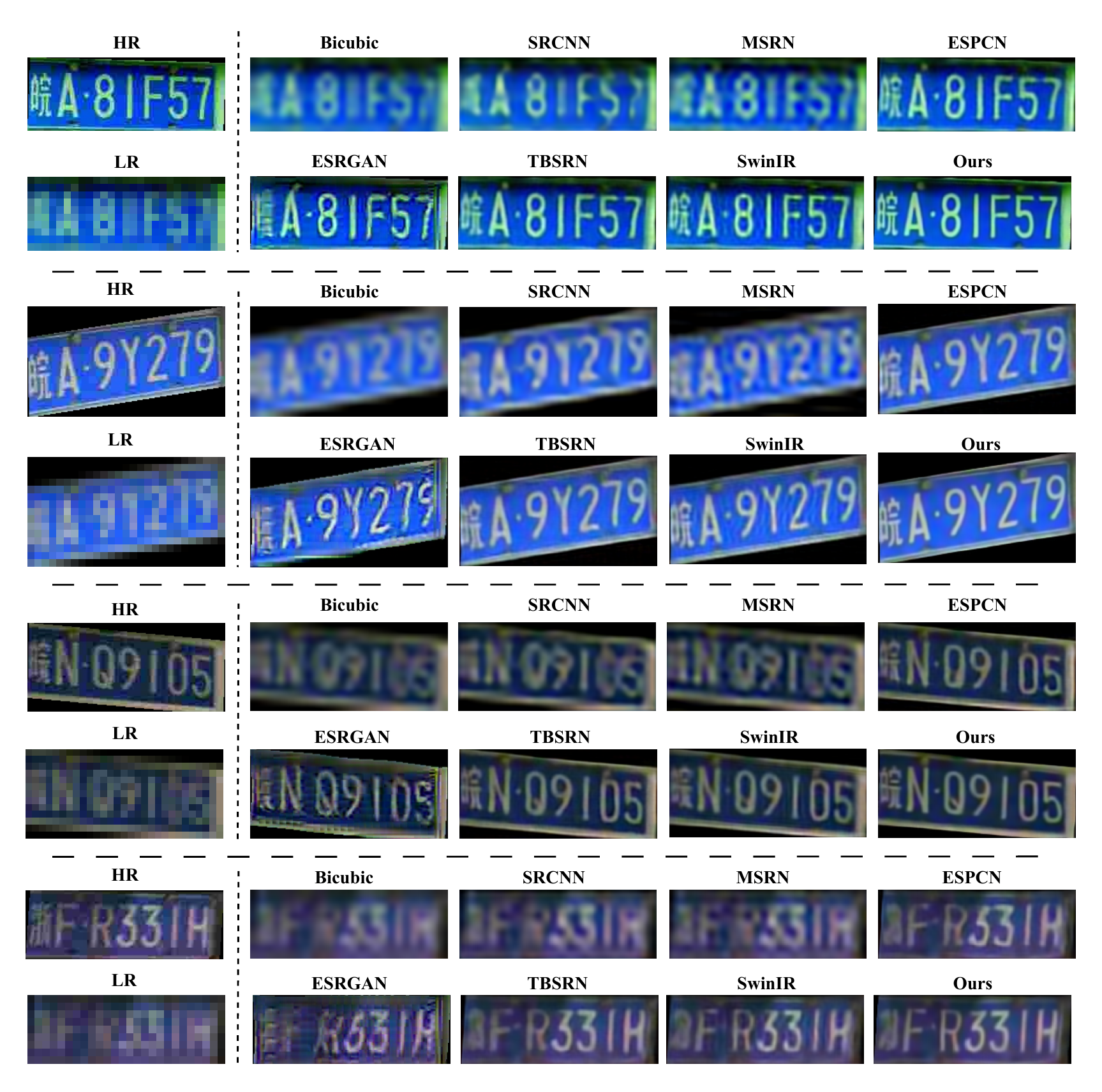}
    \caption{\textcolor{black}{Qualitative comparison with state-of-the-art methods on various samples from the CCPD dataset~\citep{xu2018towards}, taken under different conditions.}}
    \label{fig:quali_comp_sota}
\end{figure}

\begin{figure}[h]
    \centering
    \includegraphics[width=\linewidth]{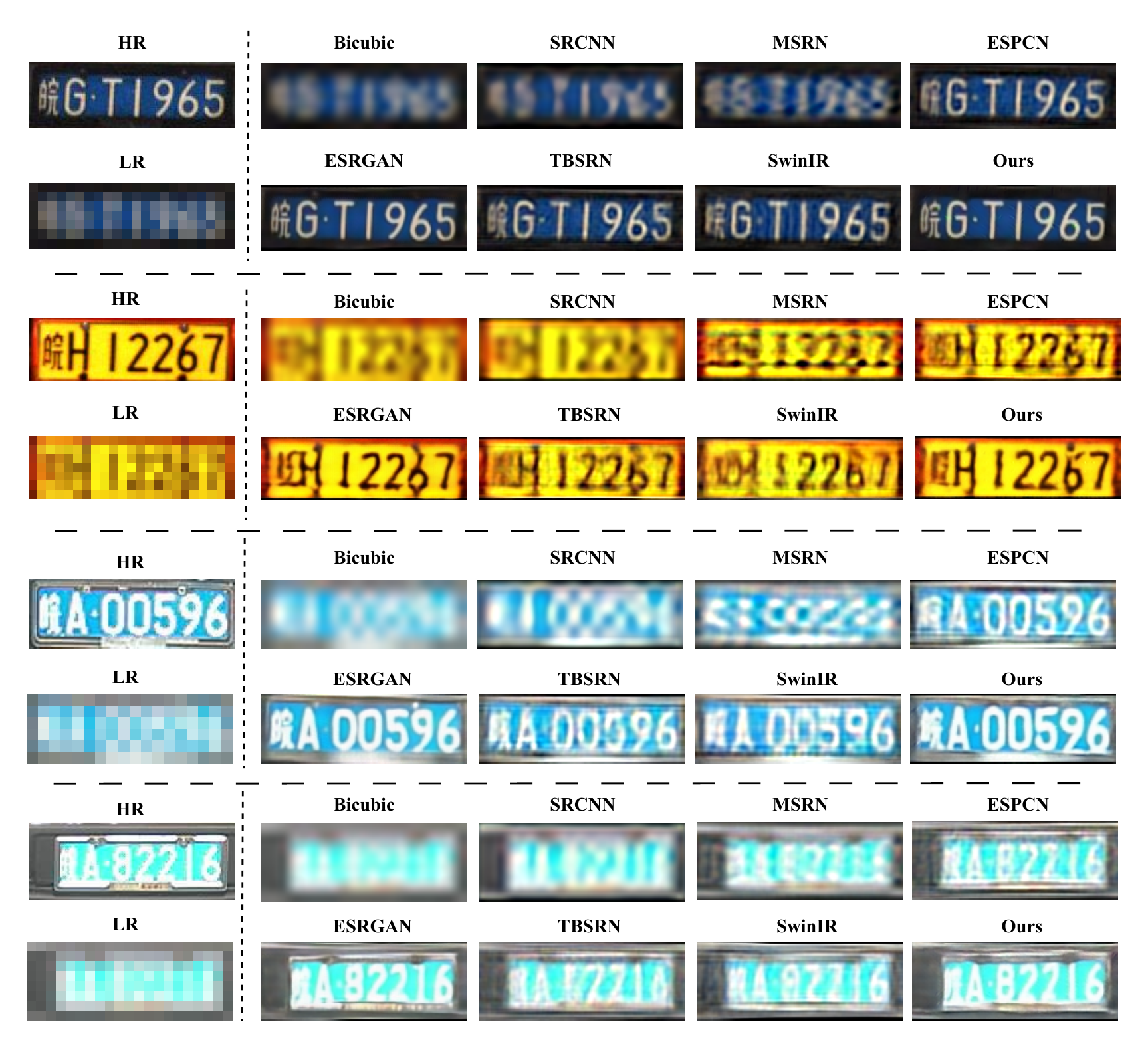}
    \caption{\textcolor{black}{Qualitative comparison with state-of-the-art methods on various samples from the PKU dataset~\citep{7752971}, taken under different conditions.}}
    \label{fig:quali_comp_sota_pku}
\end{figure}

\textbf{Qualitative comparison:}
To further compare the performance of the proposed model with state-of-the-art methods, we provide a qualitative comparison in Fig.~\ref{fig:quali_comp_sota} for CCPD and Fig.~\ref{fig:quali_comp_sota_pku} for PKU dataset. This comparison showcases the visual differences in SR outputs across various models compared to the proposed one, including Bicubic interpolation, SRCNN, MSRN, ESPCN, ESRGAN, TBSRN, and SwinIR. The results indicate significant variability in the restoration of fine details and overall visual fidelity. 

\textcolor{black}{
On the CCPD dataset, the Bicubic interpolation serves as a baseline, delivering overly smooth images with a lack of texture recovery, underscoring its inefficiency to reconstruct high-frequency details. SRCNN, being one of the earliest SR models, provides slightly sharper results but still struggles with generating realistic textures, leading to noticeable artifacts in regions with fine details. MSRN and ESPCN demonstrate improved performance, with MSRN showing better preservation of structural information and ESPCN excelling in edge sharpness. However, both models exhibit some limitations in producing natural textures. ESRGAN's outputs, known for its GAN-based approach, introduce more realistic details and textures but often at the cost of over-enhanced or unnatural artifacts in certain areas, especially on less complex regions. TBSRN and SwinIR deliver a strong balance between detail recovery and artifact suppression, with SwinIR slightly outperforming TBSRN in preserving structural consistency across diverse regions of the images. The proposed model achieves the best results overall, demonstrating superior texture restoration, edge sharpness, and fidelity to high-resolution references, with minimal artifacts and a more natural appearance.}

\textcolor{black}{
The qualitative comparison on the PKU dataset reveals the significant challenges posed by heavily degraded low-resolution inputs, where baseline methods such as Bicubic, SRCNN, and MSRN struggle to recover high-frequency details, resulting in outputs that are notably blurry and lack sharpness. ESPCN offers a modest improvement by slightly enhancing edge preservation. Methods with more advanced architectures using ViTs like TBSRN and SwinIR  further reduce blurring and improve detail recovery. Notably, ESRGAN exhibits a marked improvement on the PKU dataset compared to its performance on the CCPD dataset, a difference that can be largely attributed to the inherent characteristics of PKU, \textit{i.e.} its smaller scale and reduced variability make it more suitable to ESRGAN’s GAN-based enhancement of realistic textures. Despite this, the proposed model consistently stands out by generating images with sharp edges and rich textures that closely match the high-resolution references, again demonstrating its robust ability to reconstruct perceptually meaningful details even from severely degraded inputs.}

\textcolor{black}{
\textbf{Recognition performance analysis:} We conduct a license plate recognition analysis using several metrics as described in Sec.\ref{sec:cret} to further analyze the efficiency of the proposed method in recovering task-aware characteristics that facilitate better recognition on the CCPD dataset.  
Table~\ref{tab:text_recognition} presents a comprehensive comparison of text recognition performance on super-resolved license plate images across several methods on the CCPD datset. Bicubic interpolation, serving as a baseline, achieves an EMA of 23.70\% and an average $L_{\text{sim}}$ of 60.88\%, with high error rates (CER of 39.12\% and WER of 76.30\%), indicating its limited capacity to restore fine textual details. Early CNN-based methods like SRCNN and MSRN perform even lower, with EMA of 17.50\% and 18.30\% respectively, and similarly high error metrics, underscoring their challenges in capturing and reconstructing complex text structures. In contrast, methods leveraging deeper architectures and more advanced perceptual losses show marked improvements: ESPCN reaches an EMA of 56.60\% with a CER of 15.02\% and WER of 43.40\%, while ESRGAN, despite its GAN-based approach, records a moderate performance with 43.00\% EMA. Notably, TBSRN and SwinIR further enhance performance, achieving EMA of 57.40\% and 58.10\%, accompanied by lower error rates and improved character-level precision, recall, and F1-scores. The proposed method outperforms all competitors by reaching the highest EMA of 62.80\% and the best average $L_{\text{sim}}$ of 88.38\%, along with the lowest CER (11.62\%) and WER (37.20\%). Its superior character-level metrics, with precision, recall, and F1-scores exceeding 82\%. These performances underscore its effectiveness in enhancing textual legibility and recognition accuracy. These results collectively demonstrate that the proposed approach offers a significant advancement in recovering perceptually meaningful details from degraded license plate images, thereby enabling more accurate text recognition compared to both conventional and state-of-the-art super-resolution methods.}

\begin{table}[htbp]
\centering
\caption{Text recognition performance on super-resolved license plates from the CCPD dataset. The best and second-best results are highlighted in \textcolor{red}{\textbf{bold red}} and \textcolor{blue}{\textbf{bold blue}}, respectively.}
\label{tab:text_recognition}
\rowcolors{1}{gray!10}{white}
\setlength{\tabcolsep}{2pt}
\begin{tabular}{l|l|l|l|l|l|l|l}
\toprule
\textbf{Method} & \textbf{EMA$\uparrow$} & \textbf{L-sim$\uparrow$} & \textbf{CER$\downarrow$} & \textbf{WER$\downarrow$} & \textbf{Prec.$\uparrow$} & \textbf{Recall$\uparrow$} & \textbf{F1-S$\uparrow$} \\
\midrule
\textbf{Bicubic}  & 23.70\% & 60.88\% & 39.12\% & 76.30\% & 49.47\% & 49.72\% & 49.36\% \\
\textbf{SRCNN~\citep{dong2015image}}    & 17.50\% & 52.47\% & 47.53\% & 82.50\% & 40.37\% & 40.53\% & 40.21\%  \\
\textbf{MSRN~\citep{li2018multi}}     & 18.30\% & 55.90\% & 44.10\% & 81.70\% & 42.80\% & 43.09\% & 42.70\% \\
\textbf{ESPCN~\citep{shi2016real}}    & 56.60\% & 84.98\% & 15.02\% & 43.40\% & 78.16\% & 78.41\% & 78.11\%\\
\textbf{ESRGAN~\citep{wang2018esrgan}}   & 43.00\% & 76.00\% & 24.00\% & 57.00\% & 67.15\% & 67.52\% & 67.09\%\\
\textbf{TBSRN~\citep{chen2021scene}}    & 57.40\% & 85.67\% & 14.33\% & 42.60\% & 79.09\% & 79.22\% & 79.00\%\\
\textbf{SwinIR~\citep{liang2021swinir}}   & \textcolor{blue}{\textbf{58.10\%}} & \textcolor{blue}{\textbf{86.10\%}} & \textcolor{blue}{\textbf{13.90\%}} & \textcolor{blue}{\textbf{41.90\%}} & \textcolor{blue}{\textbf{79.63\%}} & \textcolor{blue}{\textbf{79.86\%}} & \textcolor{blue}{\textbf{79.57\%}}\\
\textbf{Ours}     & \textcolor{red}{\textbf{62.80\%}} & \textcolor{red}{\textbf{88.38\%}} & \textcolor{red}{\textbf{11.62\%}} & \textcolor{red}{\textbf{37.20\%}} & \textcolor{red}{\textbf{82.07\%}} & \textcolor{red}{\textbf{82.20\%}} & \textcolor{red}{\textbf{82.01\%}}\\
\bottomrule
\end{tabular}
\end{table}

\subsubsection{Ablation experiment}

We conduct an ablation study to verify the effectiveness of the main components of the proposed method, focusing on three key aspects: (1) the comparison between the traditional MSE loss and the proposed PECL, (2) the impact of the embedding dimensionality $d$ used for the contrastive loss, and (3) the choice of distance metrics (Euclidean vs. Manhattan) for measuring embedding similarity.

\textcolor{black}{To validate the efficiency of the proposed PECL over widely used losses, we compare it performances with solely using MSE and MAE. The performances are reported in Table~\ref{tab:comparison_losses}. As it can be seen, the proposed PECL, which combines contrastive learning on embeddings with pixel-wise reconstruction, demonstrates consistent superiority over standalone MSE and MAE losses across both the CCPD (large-scale) and PKU (smaller-scale) datasets. On the CCPD dataset, PECL achieves the highest PSNR (25.25 dB) and SSIM (0.8133), outperforming MAE by +0.26 dB and MSE by +1.03 dB in PSNR, while also attaining the best perceptual quality (LPIPS: 0.1031). This indicates that the integration of contrastive loss enhances structural fidelity and pixel-wise accuracy in large-scale settings, likely by leveraging robust embeddings to align high-resolution (HR) and super-resolved (SR) features. Notably, MAE alone outperforms MSE on CCPD, suggesting its outlier robustness benefits dense, diverse data.
}

\textcolor{black}{
On the PKU dataset, PECL again outperforms both baselines, achieving a PSNR of 19.26 dB (+0.56 dB over MSE and +0.65 dB over MAE) and the lowest LPIPS (0.1013). However, MSE slightly surpasses MAE here (PSNR: 18.70 vs. 18.61 dB), likely due to MSE’s quadratic penalty acting as a regularizer in low-data regimes, enforcing stricter pixel-level alignment critical for structured tasks like license plate super-resolution. The contrastive component of PECL mitigates MSE’s tendency to over-smooth textures, as evidenced by improved SSIM (0.7009 vs. 0.6758 for MSE).}

\begin{table*}[h]
    \setlength{\tabcolsep}{1pt} 
    \centering
    \caption{Performance comparison of the proposed PECL against baseline configurations (MSE, MAE) on the CCPD and PKU datasets. Best performances per dataset are highlighted in \textcolor{red}{\textbf{bold red}}.}
    \label{tab:comparison_losses}
    \rowcolors{0}{gray!10}{white} 
    \begin{tabular}{p{1cm}|c|c|c|c|c}
        \toprule
        \textbf{Loss} & \textbf{PSNR} & \textbf{PSNRy} & \textbf{SSIM} & \textbf{SSIMy} & \textbf{LPIPS} \\
        \midrule
        \multicolumn{6}{c}{CCPD}\\
        \midrule
        MSE & 24.22 ($\pm$\textcolor{gray}{2.34}) & 24.91 ($\pm$\textcolor{gray}{2.49}) & 0.7787 ($\pm$\textcolor{gray}{0.06}) & 0.8143 ($\pm$\textcolor{gray}{0.06}) & 0.1293 ($\pm$\textcolor{gray}{0.05}) \\
        MAE & 24.99 ($\pm$\textcolor{gray}{2.49}) & 25.78 ($\pm$\textcolor{gray}{2.64}) & 0.8118 ($\pm$\textcolor{gray}{0.06}) & 0.8428 ($\pm$\textcolor{gray}{0.06}) & 0.1090 ($\pm$\textcolor{gray}{0.04}) \\
        PECL & \textcolor{red}{\textbf{25.25}} ($\pm$\textcolor{gray}{2.61}) & \textcolor{red}{\textbf{26.12}} ($\pm$\textcolor{gray}{2.82}) & \textcolor{red}{\textbf{0.8133}} ($\pm$\textcolor{gray}{0.06}) & \textcolor{red}{\textbf{0.8461}} ($\pm$\textcolor{gray}{0.06}) & \textcolor{red}{\textbf{0.1031}} ($\pm$\textcolor{gray}{0.04}) \\
        \midrule
        \multicolumn{6}{c}{PKU}\\
        \midrule
        MSE & 18.70 ($\pm$\textcolor{gray}{3.76}) & 18.88 ($\pm$\textcolor{gray}{3.86}) & 0.6758 ($\pm$\textcolor{gray}{0.13}) & 0.6979 ($\pm$\textcolor{gray}{0.13}) & 0.1058 ($\pm$\textcolor{gray}{0.04}) \\
        MAE & 18.61 ($\pm$\textcolor{gray}{3.86}) & 18.78 ($\pm$\textcolor{gray}{3.96}) & 0.6827 ($\pm$\textcolor{gray}{0.13}) & 0.7034 ($\pm$\textcolor{gray}{0.13}) & 0.1023 ($\pm$\textcolor{gray}{0.04}) \\
        PECL & \textcolor{red}{\textbf{19.26}} ($\pm$\textcolor{gray}{3.35}) & \textcolor{red}{\textbf{19.46}} ($\pm$\textcolor{gray}{3.77}) & \textcolor{red}{\textbf{0.7009}} ($\pm$\textcolor{gray}{0.12}) & \textcolor{red}{\textbf{0.7222}} ($\pm$\textcolor{gray}{0.12}) & \textcolor{red}{\textbf{0.1013}} ($\pm$\textcolor{gray}{0.04}) \\
        \bottomrule
    \end{tabular}
\end{table*}

To evaluate the impact of embedding dimensionality and distance measures on the LPSR model performance, we conducted a comparative analysis using PSNR/PSNRy, SSIM/SSIMy, and LPIPS. The evaluation spans embedding dimensions of 64, 128, 256, and 512, trained with Euclidean and Manhattan distances. These metrics provide insights into both the fidelity and perceptual quality of the reconstructed images. The results, summarized in Table~\ref{tab:abl_pecl_ccpd} and Table~\ref{tab:abl_pecl_pku}, highlight the influence of these factors on model performance.

\begin{table*}[h]
    \setlength{\tabcolsep}{1.5pt}  
    \centering
    \caption{Performance of the proposed PECL \textit{w.r.t} the embedding distance and dimension on the CCPD dataset~\citep{xu2018towards}. The median (\textcolor{gray}{$\pm$ standard deviation}) over the testing set is reported. The best performances are highlighted in \textcolor{red}{\textbf{bold red}}.}
    \label{tab:abl_pecl_ccpd}
    \rowcolors{0}{gray!10}{white} 
    \begin{tabular}{p{1cm}|c|c|c|c|c}
        \toprule
        \textbf{Emb. dim.} & \textbf{PSNR} & \textbf{PSNRy} & \textbf{SSIM} & \textbf{SSIMy} & \textbf{LPIPS} \\
        \midrule
        \multicolumn{6}{c}{Manhattan distanc}\\
        \midrule
        64 & 25.10 ($\pm$\textcolor{gray}{2.43}) & 25.92 ($\pm$\textcolor{gray}{2.58}) & 0.8126 ($\pm$\textcolor{gray}{0.06}) & 0.8460 ($\pm$\textcolor{gray}{0.06}) & 0.1075 ($\pm$\textcolor{gray}{0.04}) \\
        128 & \textcolor{red}{\textbf{25.25}} ($\pm$\textcolor{gray}{2.61}) & \textcolor{red}{\textbf{26.12}} ($\pm$\textcolor{gray}{2.82}) & \textcolor{red}{\textbf{0.8133}} ($\pm$\textcolor{gray}{0.05}) & \textcolor{red}{\textbf{0.8461}} ($\pm$\textcolor{gray}{0.06}) & \textcolor{red}{\textbf{0.1031}} ($\pm$\textcolor{gray}{0.03}) \\
        256 & 25.11 ($\pm$\textcolor{gray}{2.43}) & 25.92 ($\pm$\textcolor{gray}{2.58}) & 0.8126 ($\pm$\textcolor{gray}{0.06}) & 0.8460 ($\pm$\textcolor{gray}{0.06}) & 0.1075 ($\pm$\textcolor{gray}{0.04}) \\
        512 & 25.12 ($\pm$\textcolor{gray}{2.46}) & 25.93 ($\pm$\textcolor{gray}{2.61}) & 0.8127 ($\pm$\textcolor{gray}{0.06}) & 0.8458 ($\pm$\textcolor{gray}{0.06}) & 0.1065 ($\pm$\textcolor{gray}{0.04}) \\
        \midrule
        \rowcolor{white}
        \multicolumn{6}{c}{Euclidean distanc}\\
        \midrule
        64 & 24.22 ($\pm$\textcolor{gray}{2.38}) & 24.89 ($\pm$\textcolor{gray}{2.52}) & 0.7511 ($\pm$\textcolor{gray}{0.05}) & 0.7841 ($\pm$\textcolor{gray}{0.06}) & 0.1005 ($\pm$\textcolor{gray}{0.04}) \\
        128 & 24.37 ($\pm$\textcolor{gray}{2.40}) & 25.06 ($\pm$\textcolor{gray}{2.53}) & 0.7562 ($\pm$\textcolor{gray}{0.07}) & 0.7875 ($\pm$\textcolor{gray}{0.06}) & 0.0959 ($\pm$\textcolor{gray}{0.04}) \\
        256 & 24.54 ($\pm$\textcolor{gray}{2.40}) & 25.24 ($\pm$\textcolor{gray}{2.53}) & 0.7670 ($\pm$\textcolor{gray}{0.05}) & 0.7980 ($\pm$\textcolor{gray}{0.05}) & 0.0926 ($\pm$\textcolor{gray}{0.03}) \\
        512 & 24.58 ($\pm$\textcolor{gray}{2.45}) & 25.29 ($\pm$\textcolor{gray}{2.58}) & 0.7715 ($\pm$\textcolor{gray}{0.05}) & 0.8013 ($\pm$\textcolor{gray}{0.06}) & 0.0948 ($\pm$\textcolor{gray}{0.03}) \\
        \bottomrule
    \end{tabular}
\end{table*}

\begin{table*}[h]
    \setlength{\tabcolsep}{1.5pt} 
    \centering
    \caption{Performance of the proposed PECL \textit{w.r.t} the embedding distance and dimension on the PKU dataset~\citep{7752971}. The median (\textcolor{gray}{$\pm$ standard deviation}) over the testing set is reported. The best performances are highlighted in \textcolor{red}{\textbf{bold red}}.}
    \label{tab:abl_pecl_pku}
    \rowcolors{0}{gray!10}{white} 
    \begin{tabular}{p{1cm}|c|c|c|c|c}
        \toprule
        \textbf{Emb. dim.} & \textbf{PSNR} & \textbf{PSNRy} & \textbf{SSIM} & \textbf{SSIMy} & \textbf{LPIPS} \\
        \midrule
        \multicolumn{6}{c}{Manhattan distance}\\
        \midrule
        64 & 18.84 ($\pm$\textcolor{gray}{3.83}) & 19.02 ($\pm$\textcolor{gray}{3.94}) & 0.6821 ($\pm$\textcolor{gray}{0.13}) & 0.7032 ($\pm$\textcolor{gray}{0.14}) & 0.1017 ($\pm$\textcolor{gray}{0.04}) \\
        128 & 19.16 ($\pm$\textcolor{gray}{3.85}) & 19.34 ($\pm$\textcolor{gray}{3.98}) & 0.6909 ($\pm$\textcolor{gray}{0.12}) & 0.7124 ($\pm$\textcolor{gray}{0.13}) & 0.1013 ($\pm$\textcolor{gray}{0.04}) \\
        256 & 18.61 ($\pm$\textcolor{gray}{3.80}) & 18.79 ($\pm$\textcolor{gray}{3.92}) & 0.6796 ($\pm$\textcolor{gray}{0.13}) & 0.7005 ($\pm$\textcolor{gray}{0.13}) & 0.1006 ($\pm$\textcolor{gray}{0.04}) \\
        512 & 18.90 ($\pm$\textcolor{gray}{3.89}) & 19.09 ($\pm$\textcolor{gray}{4.00}) & 0.6905 ($\pm$\textcolor{gray}{0.12}) & 0.7114 ($\pm$\textcolor{gray}{0.13}) & 0.0999 ($\pm$\textcolor{gray}{0.04}) \\
        \midrule
        \rowcolor{white}
        \multicolumn{6}{c}{Manhattan distance}\\
        \midrule
        64 & \textcolor{red}{\textbf{19.26}} ($\pm$\textcolor{gray}{3.65}) & \textcolor{red}{\textbf{19.46}} ($\pm$\textcolor{gray}{3.77}) & \textcolor{red}{\textbf{0.7009}} ($\pm$\textcolor{gray}{0.12}) & \textcolor{red}{\textbf{0.7222}} ($\pm$\textcolor{gray}{0.12}) & 0.1013 ($\pm$\textcolor{gray}{0.04}) \\
        128 & 19.15 ($\pm$\textcolor{gray}{3.65}) & 19.34 ($\pm$\textcolor{gray}{3.75}) & 0.6933 ($\pm$\textcolor{gray}{0.12}) & 0.7147 ($\pm$\textcolor{gray}{0.12}) & 0.1001 ($\pm$\textcolor{gray}{0.04}) \\
        256 & 18.96 ($\pm$\textcolor{gray}{3.63}) & 19.14 ($\pm$\textcolor{gray}{3.74}) & 0.6907 ($\pm$\textcolor{gray}{0.12}) & 0.7119 ($\pm$\textcolor{gray}{0.12}) & 0.0990 ($\pm$\textcolor{gray}{0.04}) \\
        512 & 19.08 ($\pm$\textcolor{gray}{3.68}) & 19.28 ($\pm$\textcolor{gray}{3.81}) & 0.6923 ($\pm$\textcolor{gray}{0.12}) & 0.7136 ($\pm$\textcolor{gray}{0.12}) & \textcolor{red}{\textbf{0.0987}} ($\pm$\textcolor{gray}{0.04}) \\
        \bottomrule
    \end{tabular}
\end{table*}

\textcolor{black}{
The ablation study on the CCPD dataset, with more than 100k training license plates, demonstrates that the Manhattan-based embedding distance outperforms Euclidean in terms of pixel wise (PSNR) and structural similarity (SSIM) quality metrics, achieving optimal performance at 128 dimensions (PSNR: 25.25 dB, SSIM: 0.8133). This superiority can be attributed to the distance's robustness to outliers and its ability to preserve fine structural details in high-data regimes, where the large and diverse training set mitigates overfitting risks. The quadratic penalty of MSE, while effective at minimizing large pixel errors, may over-penalize subtle but perceptually acceptable deviations in noisy or variable inputs, such as skewed license plates or lighting variation. This is particularly are prevalent in real-world datasets like CCPD. The contrastive loss component of PECL synergizes with Manhattan distance by learning discriminative embeddings that emphasize structural fidelity, leveraging the dataset’s scale to generalize across diverse samples. The marginal gains beyond 128 dimensions suggest a balance between embedding complexity and reconstruction accuracy, as higher dimensions risk overfitting even in large datasets.}

\textcolor{black}{
In contrast, the PKU dataset, with 3.5K training license plates, exhibits superior performance with the Euclidean-based embedding distance loss, particularly at 64 dimensions (PSNR: 19.26 dB, SSIM: 0.7009). This reversal highlights the role of dataset size and task-specific regularization. The smaller training set amplifies the Euclidean distance advantages. In particular, its quadratic penalty acts as an implicit regularizer, reducing the risk of overfitting by aggressively minimizing large pixel errors that could otherwise propagate into the embedding space. This is critical for license plate super-resolution, where rigid, high-contrast text demands pixel-perfect alignment to ensure readability. Manhattan distance’s tolerance for moderate errors, while beneficial in large datasets, may fail to enforce the geometric precision required in low-data regimes. Furthermore, the contrastive loss in PECL benefits from Euclidean distance rigor, which stabilizes embedding learning when training samples are limited. The best perceptual quality (LPIPS: 0.0987) at 512 dimensions for Euclidean distance suggests that higher embedding dimensions can enhance perceptual alignment in small datasets, but the minimal gains at lower dimensions (64) underscore the importance of computational efficiency. These results emphasize the task-data interplay, as the Euclidean distance sensitivity to subtle errors aligns with the structured nature of license plates, while its regularization effect compensates for the limited data diversity in  PKU dataset.}

\begin{equation}\label{eq:contrast}
    \text{Contrast} (\text{PECL}, \text{Base}) = \frac{\text{PSNR}_{\text{PECL}} - \text{PSNR}_{\text{Base}}}{\text{PSNR}_{\text{Base}}}.
\end{equation}

\begin{figure}[h]
    \setlength{\tabcolsep}{0pt}
    \centering
    \begin{tabular}{cc}
             \includegraphics[width=0.5\linewidth]{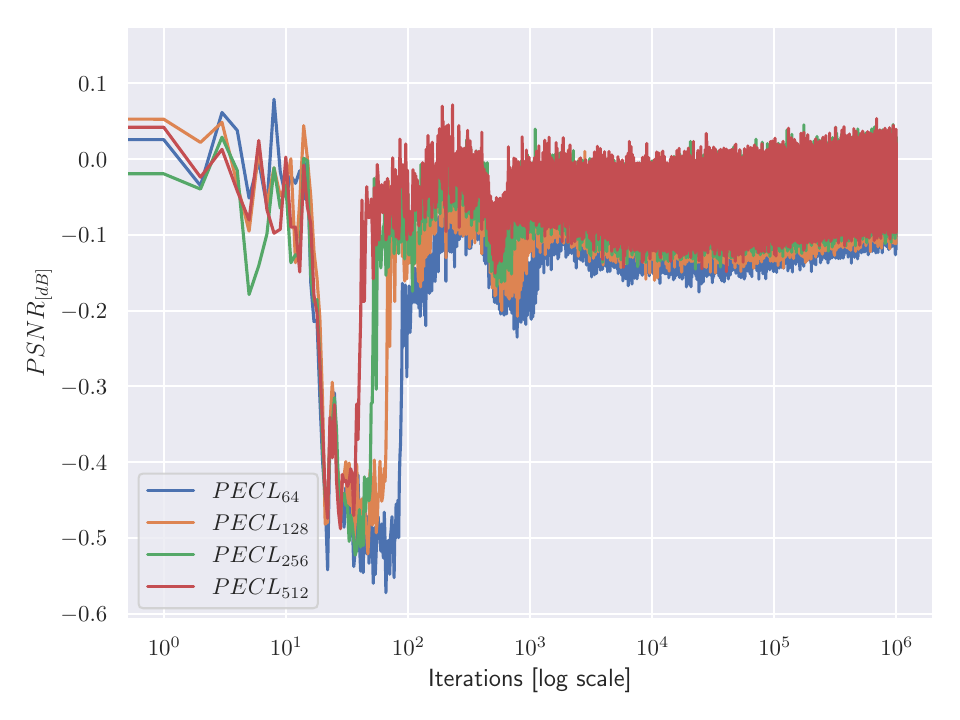}&
             \includegraphics[width=0.5\linewidth]{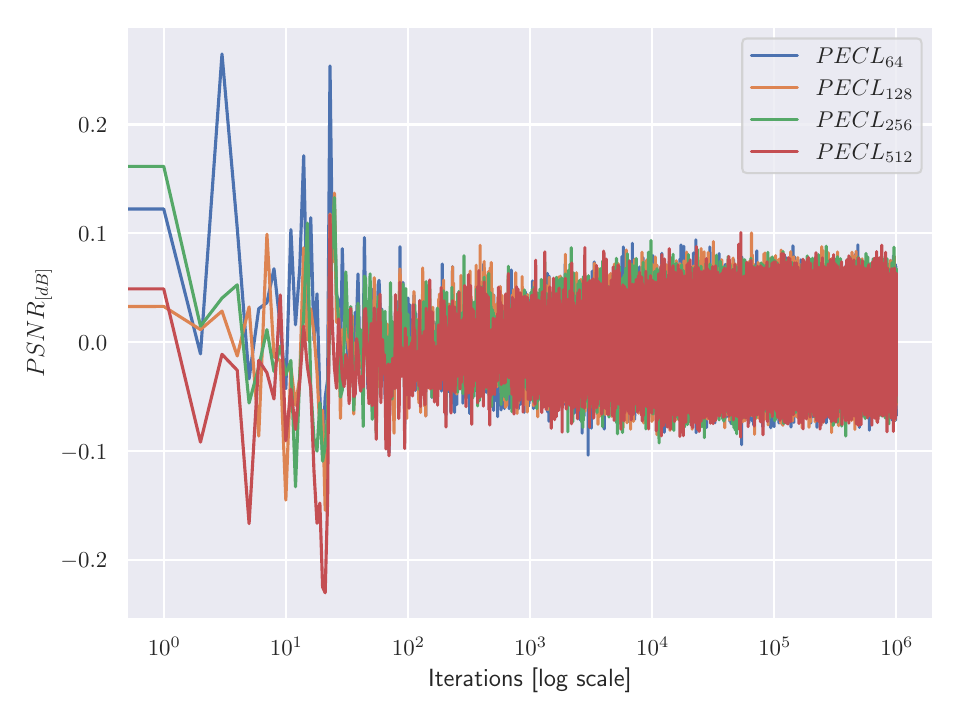}\\
    \end{tabular}
    \caption{\textcolor{black}{PECL and MSE loss relative contrast on the CCPD dataset~\citep{xu2018towards} of PSNR values with embedding sizes of 64, 128, 256, and 512, using Euclidean distance (left) and Manhattan distance (right) during training. The contrast, defined as the relative improvement in PSNR, is plotted against the training iterations on a logarithmic scale. Positive contrast values indicate better performance of the PECL models compared to MSE, while negative values indicate worse performance.}}
    \label{fig:contrast_psnr_ccpd_mse}
\end{figure}

To evaluate the training dynamics of the proposed PECL, we compare its PSNR performance to the baseline MSE and MAE losses across various embedding sizes (64, 128, 256, and 512). The analysis aims to quantify the relative improvement introduced by PECL using the contrast metric defined in Eq.\ref{eq:contrast}. This metric normalizes the difference in PSNR values by the baseline, providing a more interpretable measure of performance gains during training. The results are depicted in Fig.~\ref{fig:contrast_psnr_ccpd_mse} and Fig.~\ref{fig:contrast_psnr_ccpd_mse}, for the contrast with MSE and MAE, respectively. These curves offer insights into the learning behavior of the proposed PECL with Euclidean and Manhattan distances as embedding similarity measures. The x-axis, presented on a logarithmic scale, captures training iterations, while the y-axis shows the contrast in PSNR values (in dB).  Positive contrast values indicate that the proposed PECL outperforms the baseline, while negative values suggest underperformance. The baseline versions, MSE and MAE, favors high PSNR due to their pixel-wise loss objective. In contrast, the PECL combines pixel-wise with a contrastive loss, which focuses on embedding similarity and perceptual quality, making it less focused on maximizing PSNR directly.

\begin{figure}[h]
    \setlength{\tabcolsep}{0pt}
    \centering
    \begin{tabular}{cc}
             \includegraphics[width=0.5\linewidth]{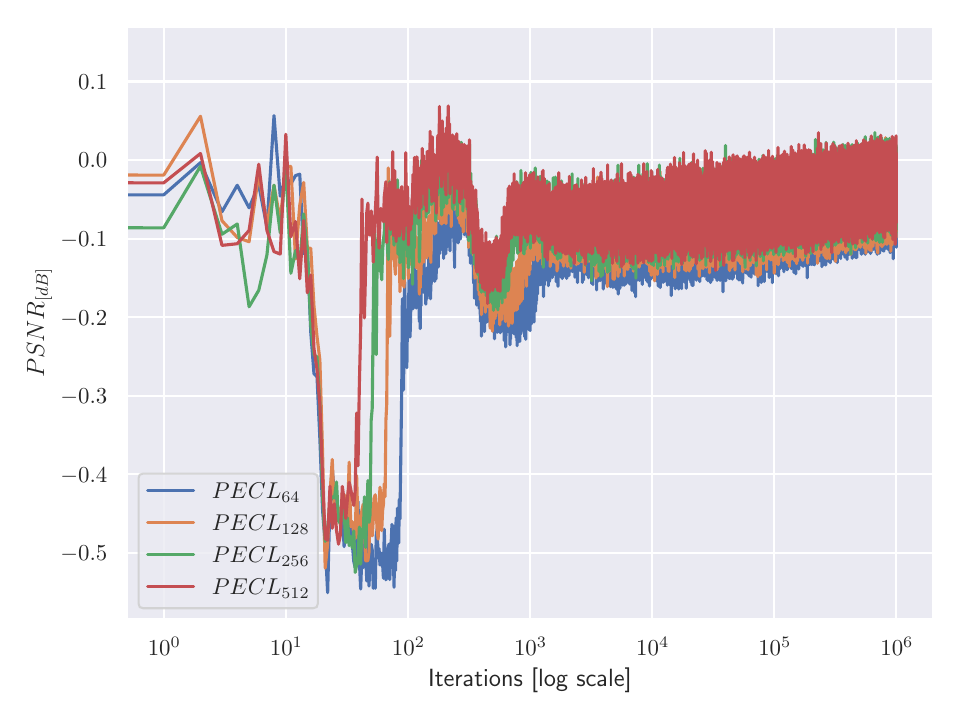}&
             \includegraphics[width=0.5\linewidth]{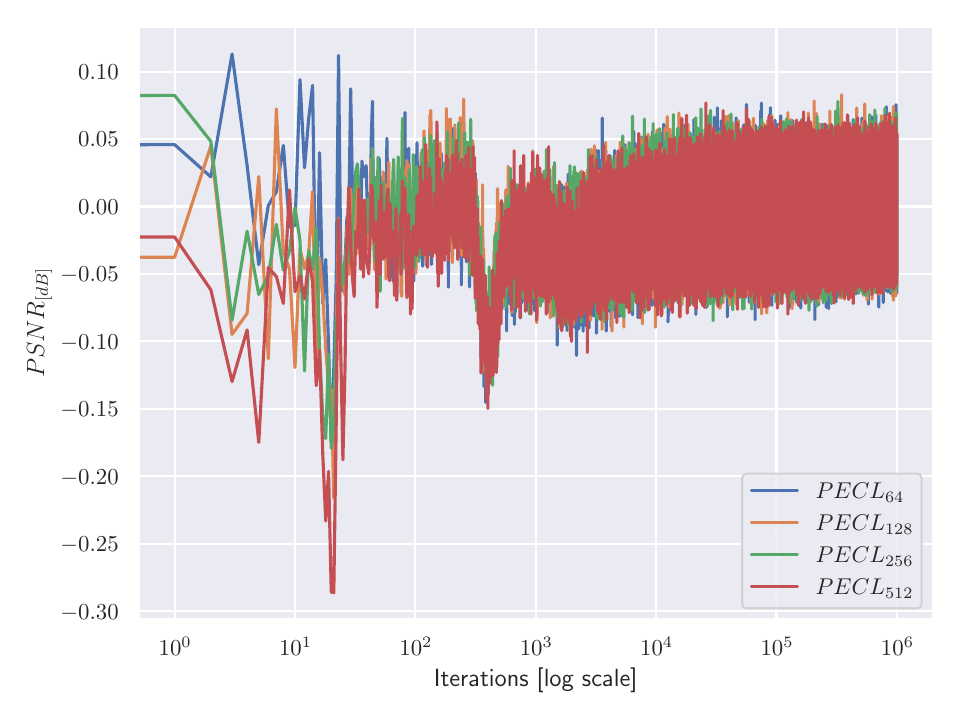}\\
    \end{tabular}
    \caption{\textcolor{black}{PECL and MAE loss relative contrast CCPD dataset~\citep{xu2018towards} of PSNR values with embedding sizes of 64, 128, 256, and 512, using Euclidean distance (left) and Manhattan distance (right) during training. The contrast, defined as the relative improvement in PSNR, is plotted against the training iterations on a logarithmic scale. Positive contrast values indicate better performance of the PECL models compared to MSE, while negative values indicate worse performance.}}
    \label{fig:contrast_psnr_ccpd_mae}
\end{figure}

\textcolor{black}{
Analyzing the contrast curves obtained on the CCPD dataset in Fig.~\ref{fig:contrast_psnr_ccpd_mse} and Fig.~\ref{fig:contrast_psnr_ccpd_mae}, the curves obtained from training the PECL with Manhattan distance are more centered around the zero axis, with values fluctuating between -0.22 and 0.25 when compared to MSE loss, and between -0.3 and 0.1 when compared to MAE loss. In contrast, the Euclidean distance curves exhibit a wider spread, ranging from -0.6 to 0.13 and -0.5 to 0.13, compared to MSE and MAE loss, respectively. This difference highlights that, while both distances show a tendency for improvement as training progresses, the Euclidean curves are more influenced by negative contrast values, indicating that Euclidean-based PECL start off with a larger disparity in performance. The more centered Manhattan curves suggest that its impact on the PSNR performance is more stable and balanced throughout training.}

\begin{figure}[htbp]
    \setlength{\tabcolsep}{0pt}
    \centering
    \begin{tabular}{cc}
             \includegraphics[width=0.5\linewidth]{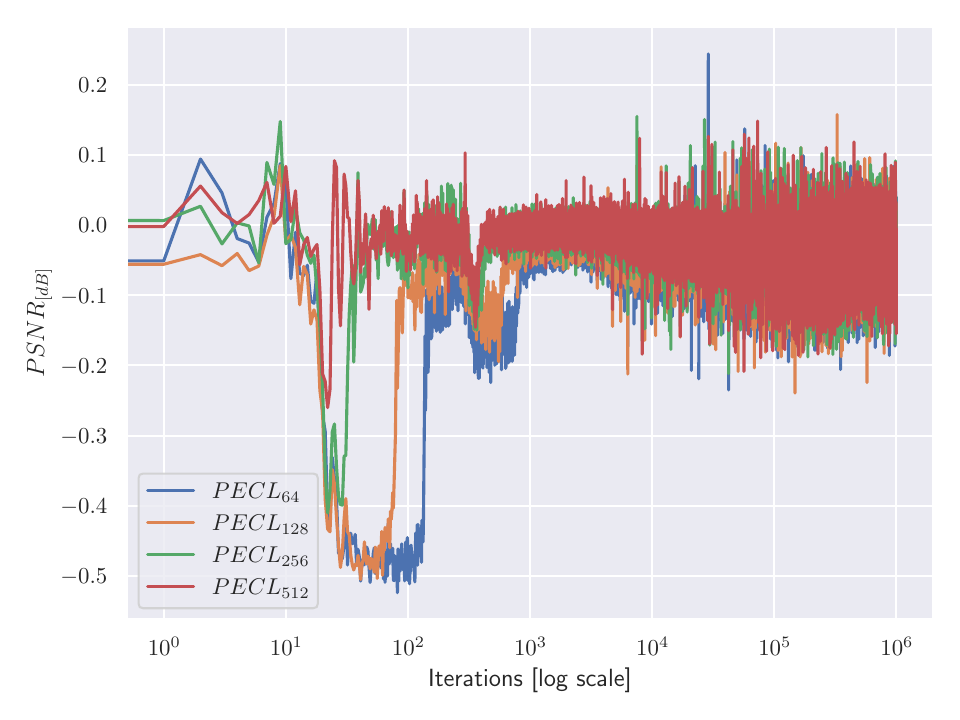}&
             \includegraphics[width=0.5\linewidth]{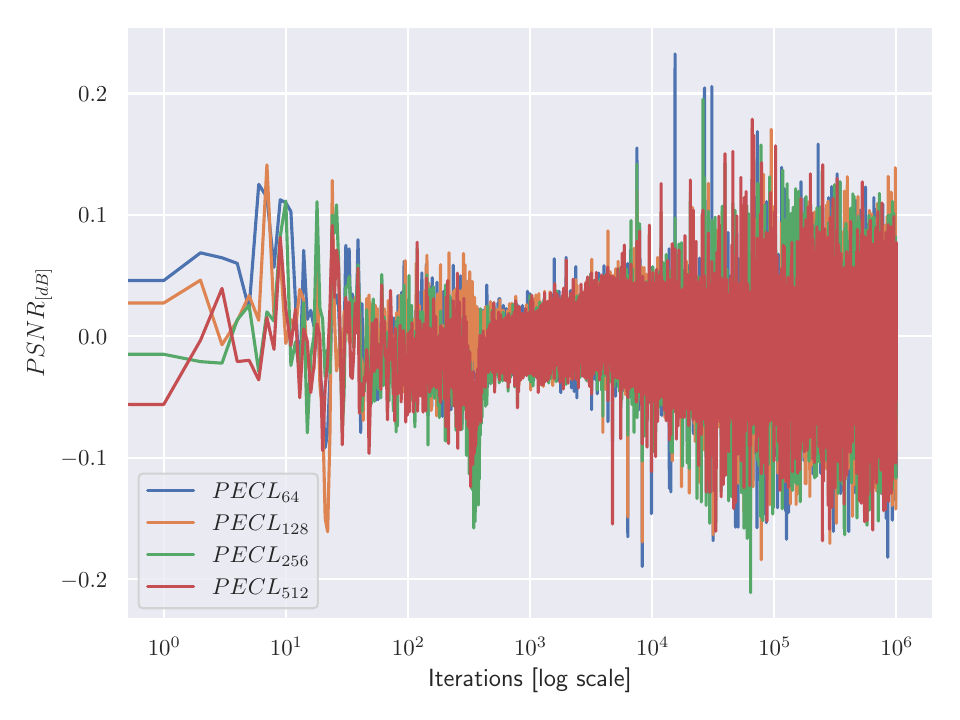}\\
    \end{tabular}
    \caption{\textcolor{black}{PECL and MSE loss relative contrast on the PKU dataset~\citep{7752971} of PSNR values with embedding sizes of 64, 128, 256, and 512, using Euclidean distance (left) and Manhattan distance (right) during training. The contrast, defined as the relative improvement in PSNR, is plotted against the training iterations on a logarithmic scale. Positive contrast values indicate better performance of the PECL models compared to MSE, while negative values indicate worse performance.}}
    \label{fig:contrast_psnr_pku_mse}
\end{figure}

\begin{figure}[htbp]
    \setlength{\tabcolsep}{0pt}
    \centering
    \begin{tabular}{cc}
             \includegraphics[width=0.5\linewidth]{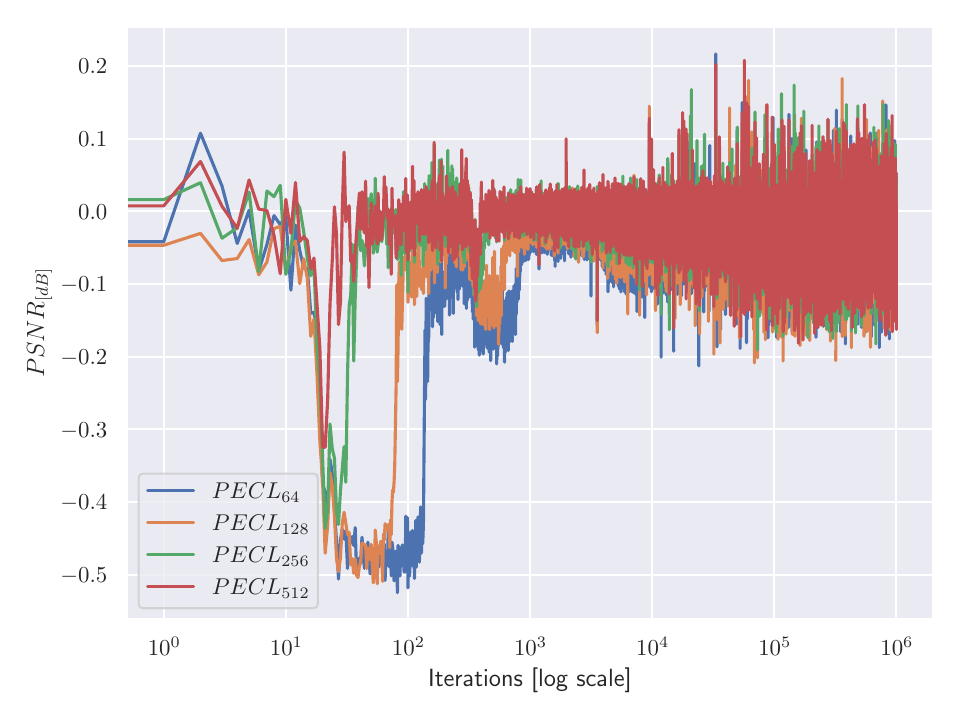}&
             \includegraphics[width=0.5\linewidth]{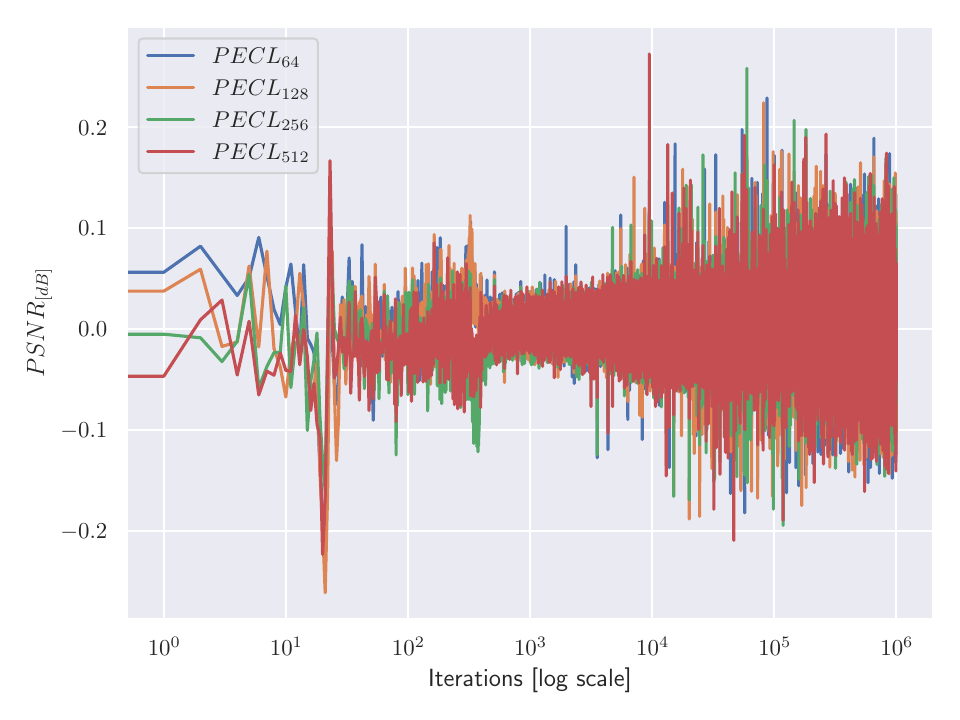}\\
    \end{tabular}
    \caption{\textcolor{black}{PECL and MAE loss relative contrast on the PKU dataset~\citep{7752971} of PSNR values with embedding sizes of 64, 128, 256, and 512, using Euclidean distance (left) and Manhattan distance (right) during training. The contrast, defined as the relative improvement in PSNR, is plotted against the training iterations on a logarithmic scale. Positive contrast values indicate better performance of the PECL models compared to MSE, while negative values indicate worse performance.}}
    \label{fig:contrast_psnr_pku_mae}
\end{figure}

Similarly, the curves obtained on the PKU dataset, depicted in Fig.~\ref{fig:contrast_psnr_pku_mse} and Fig.~\ref{fig:contrast_psnr_pku_mae} showcase the same behavior. During the initial training phase, the contrast values for both Euclidean and Manhattan exhibit fluctuations, reflecting the combined influence of MSE and contrastive loss. As training progresses, the curves stabilize and shift toward positive values, consistently exceeding zero and converging around 0.1 and 0.2 on the CCPD and PKu datset, respectively. This indicates that PECL, despite improving both pixel-wise and embedding-based similarities, achieves a higher PSNR than MSE or MAE. This reflects its ability to produce better visual quality. The positive contrast suggests that the contrastive loss contributes to improving perceptual performance, making the images visually more accurate while still maintaining relatively similar embeddings.

Larger embedding dimensionality exhibits smoother and more stable contrast curves, regardless of which baseline loss is compared to. Particularly for Manhattan, suggesting that higher-dimensional embeddings contribute to more robust feature representation and performance stability throughout training. This highlights the role of embedding dimensionality in learning distinctive features, which enhances the model's ability to generalize. The observed trends emphasize the trade-off managed by PECL, effectively balancing pixel-wise accuracy with perceptual quality to achieve better overall performance compared to the baseline MSE.

\begin{figure}[htbp]
    \centering
    \includegraphics[width=\linewidth]{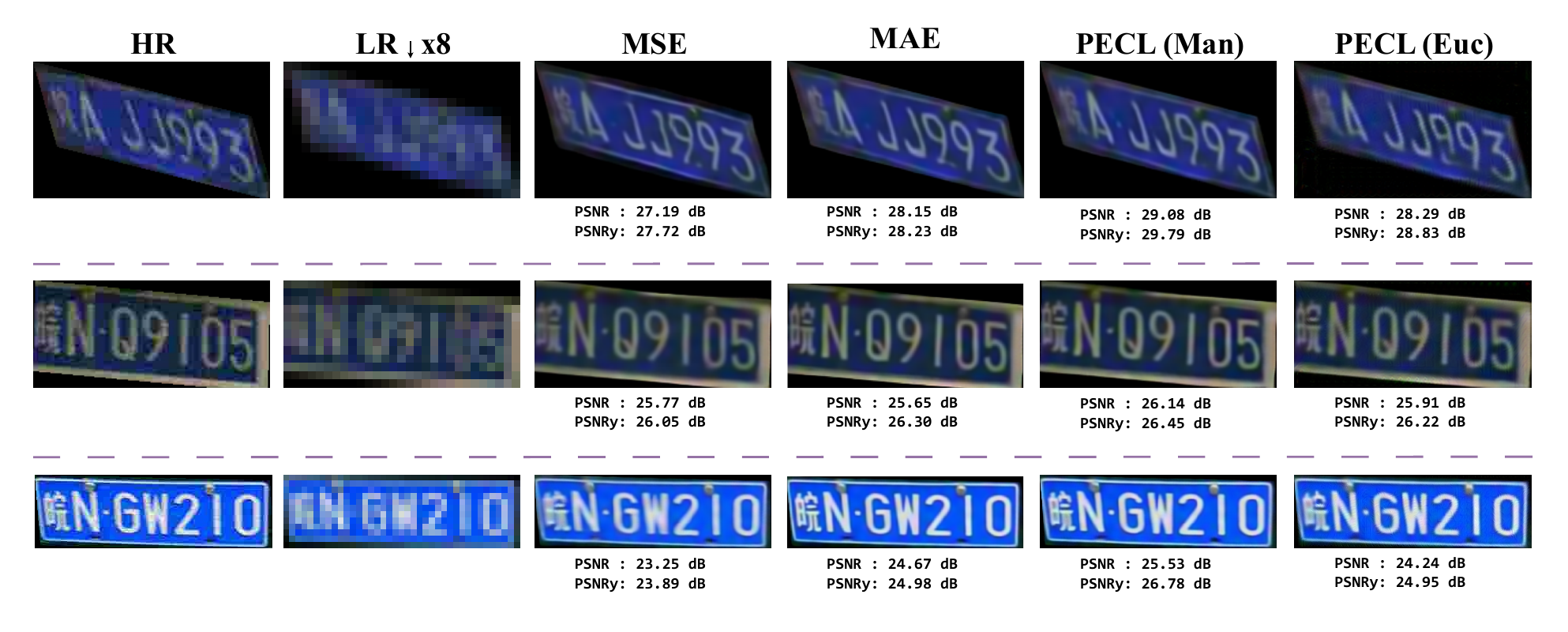}
    \caption{Qualitative comparison of license plates from the CCPD dataset~\citep{xu2018towards} under varying acquisition conditions using the proposed PECL with Manhattan and Euclidean distances, compared to the baseline MSE and MAE losses.}
    \label{fig:qualitative_comparison_ccpd}
\end{figure}

\begin{figure}[htbp]
    \centering
    \includegraphics[width=\linewidth]{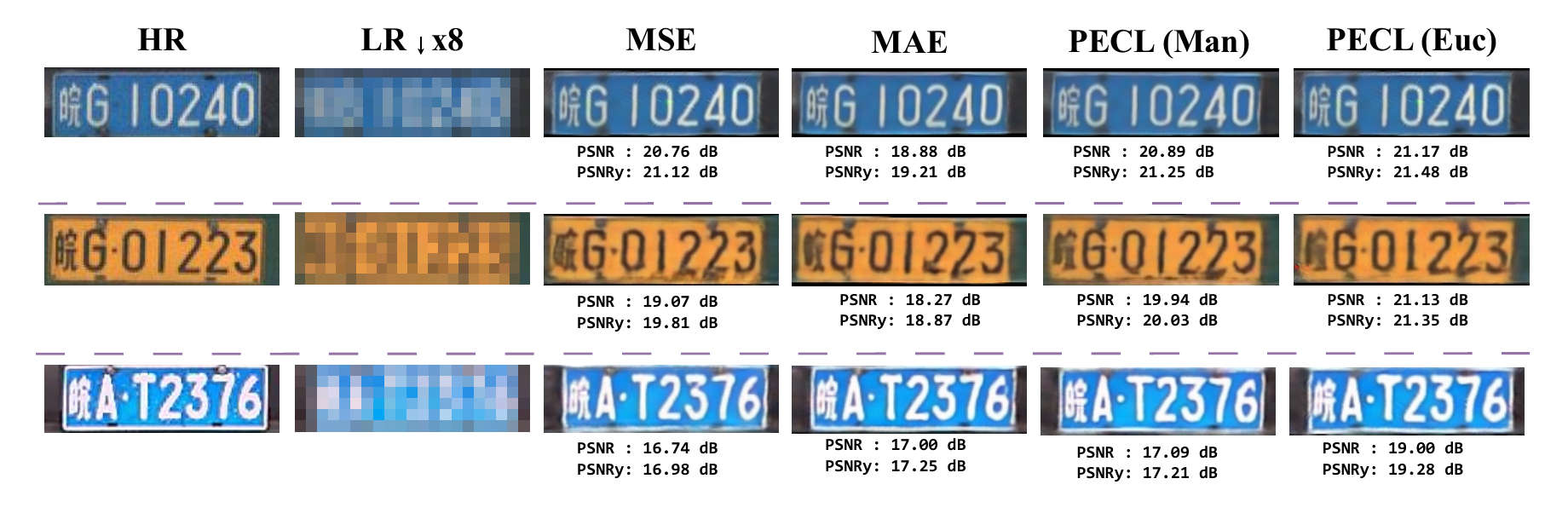}
    \caption{\textcolor{black}{Qualitative comparison of license plates from the PKU dataset~\citep{7752971} under varying acquisition conditions using the proposed PECL with Manhattan and Euclidean distances, compared to the baseline MSE and MAE losses.}}
    \label{fig:qualitative_comparison_pku}
\end{figure}

\textcolor{black}{
The qualitative comparison, as shown in Fig.~\ref{fig:qualitative_comparison_ccpd} and Fig.~\ref{fig:qualitative_comparison_pku}, provides a visual assessment of the performance of the proposed PECL framework with both Manhattan and Euclidean distances, benchmarked against the baseline MSE and MAE losses. The figures present the high-resolution (HR) reference image alongside the low-resolution (LR) input, downsized by a factor of 8, and the super-resolved (SR) results obtained using MSE, MAE, PECL (Man), and PECL (Euc). This layout allows for a direct evaluation of the visual quality achieved by each approach, highlighting the differences in detail preservation, texture sharpness, and perceptual fidelity. As can be seen, the HR images exhibit varying acquisition conditions, including differences in lighting and angles. Additionally, their visual quality appears degraded, with noticeable noise and artifacts. After downsizing them by a factor of 8 to produce the LR inputs, the fine details, such as text and edges, become significantly unclear and, in some cases, unreadable. This highlights the challenge of reconstructing accurate and perceptually meaningful SR outputs.}

\textcolor{black}{
From the CCPD dataset, PECL (Man) clearly outperforms the alternatives: in the first example, it achieves a PSNR of 29.08 dB and PSNRy of 29.79 dB, producing significantly sharper edges and more detailed textures than PECL (Euc), MSE, or MAE, which tend to smooth or blur high-frequency details such as text. In subsequent examples, PECL (Man) maintains this advantage, with PSNR values of 26.14 dB and 25.53 dB (with corresponding PSNRy improvements), thereby preserving the fine structural details evident in the high-resolution references. Conversely, for samples drawn from the PKU dataset, PECL (Euc) consistently outperforms the other losses, attaining a PSNR of 21.17 dB and PSNRy of 21.48 dB in the first sample by delivering visually sharper reconstructions and more refined textures compared to MSE, MAE, and even PECL (Man). These observations collectively underscore the benefit of integrating embedding consistency through contrastive loss: while PECL (Euc) enhances structural fidelity under certain conditions, the embedding alignment enforced by PECL (Man) appears particularly adept at recovering fine-grained details and ensuring perceptual quality, especially in high-frequency regions.}

\textcolor{black}{
It is also important to highlight the quality differences between the datasets. The CCPD dataset is large-scale, encompassing a wide variety of imaging conditions, which facilitates a more robust evaluation of super-resolution methods. On the other hand, the PKU dataset is small-scale, characterized by a limited number of images and generally lower inherent quality. This distinction emphasizes the adaptability of the proposed approach, where PECL (Man) proves particularly effective on the more varied and higher-quality CCPD dataset, while PECL (Euc) shows relative advantages on the smaller-scale, more challenging PKU dataset.}

\begin{figure}[htbp]
    \centering
    \begin{tabular}{c}
         \includegraphics[width=\linewidth]{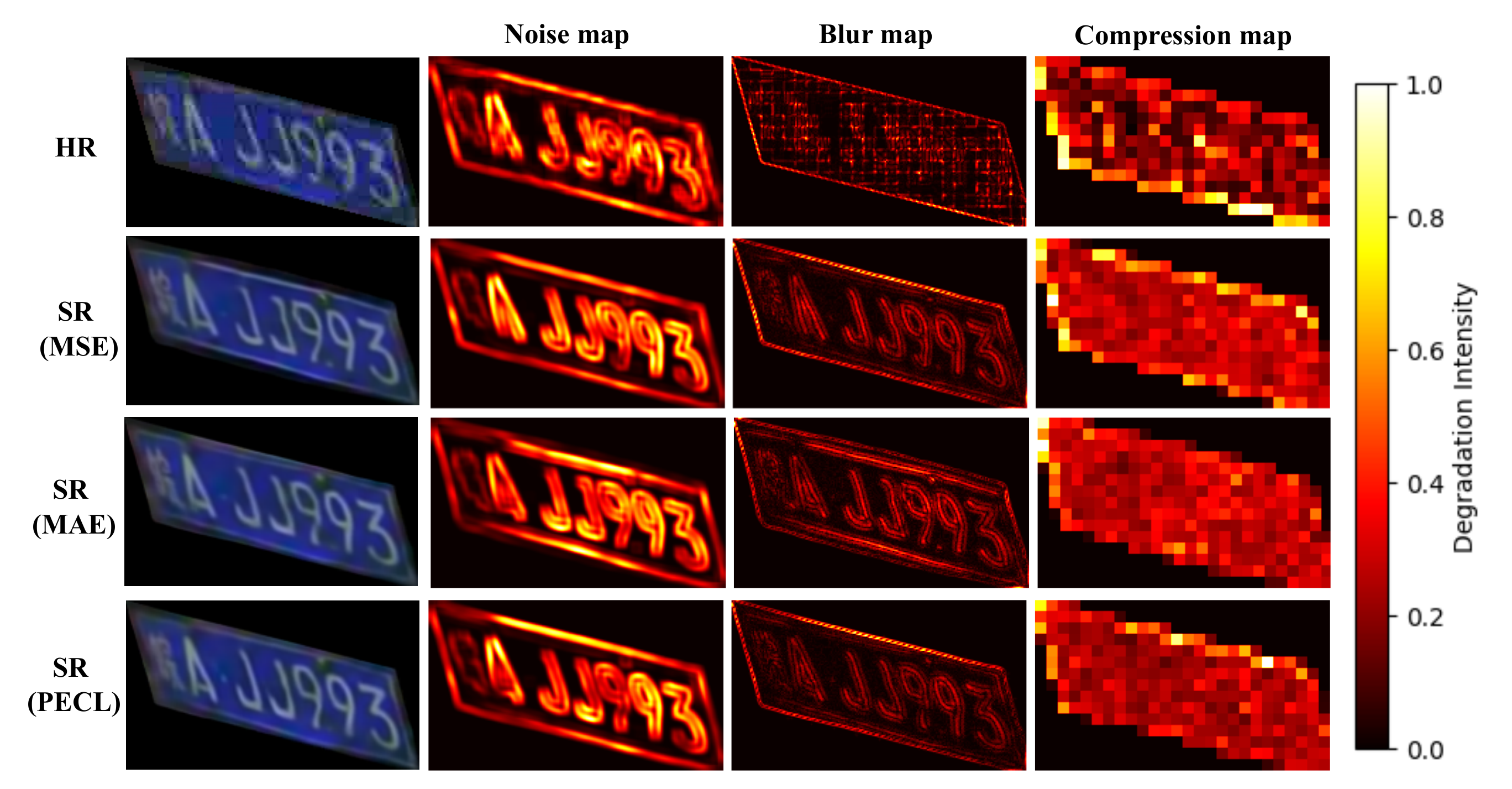}\\
         \includegraphics[width=\linewidth]{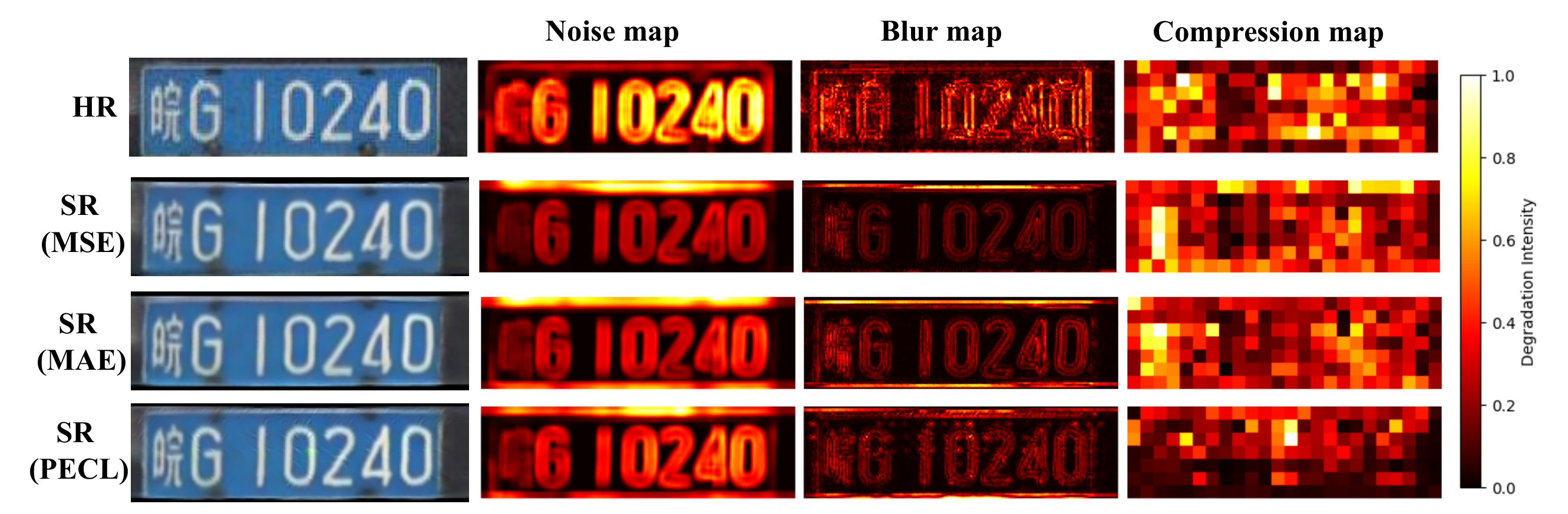}
    \end{tabular}
    \caption{Qualitative comparison of HR and SR license plate images, with degradation maps showing noise, blur, and compression artifacts. (top) CCPD dataset~\citep{xu2018towards} and (bottom) PKU dataset~\citep{7752971}. The maps illustrate the intensity of degradation in both the HR and SR images, with each map providing insights into how the visual quality is affected by the SR model trained with the proposed PECL over MSE and MAE losses.}

    \label{fig:dist_maps}
\end{figure}

\textcolor{black}{By analyzing the distortion maps (Noise, Blur, and Compression) in Fig.~\ref{fig:dist_maps} we can highlight the differences in performance when training the model with MSE or MAE compared to PECL,  with respect to specific degradations. The blur map for PECL demonstrates significantly reduced blur, regardless on which dataset, indicating that PECL can restore sharper edges and finer details. In contrast, with MSE and MAE, higher blur intensity can be observed around high-frequency regions such as edges. This reflects its tendency to produce over-smoothed outputs. The noise map reveals that PECL maintains moderate and localized noise levels, comparable to MSE. The compression map further demonstrates that PECL introduces fewer compression artifacts, particularly in textured regions, resulting in perceptually superior reconstructions. Compared to the HR image, the proposed method appears to smooth the images, resulting in distribution depicted with the blur maps, regardless of the training loss.} 

Overall, the PECL model balances sharpness, controlled noise, and minimal compression artifacts, demonstrating its effectiveness in producing detailed and high-quality SR images while avoiding significant distortions. This highlights the strength of the embedding consistency enforced by the PECL loss, which aligns features and enhances sharpness generation without compromising overall image quality.

\textcolor{black}{
\textbf{Embedding space visualization:}
We conduct a feature visualization using t-SNE visualizations of feature embeddings among Euclidean and Manhattan distances as embedding similarity measures. }

\begin{figure}[htbp]
    \centering
    \begin{tabular}{cc}
       \includegraphics[width=0.5\linewidth]{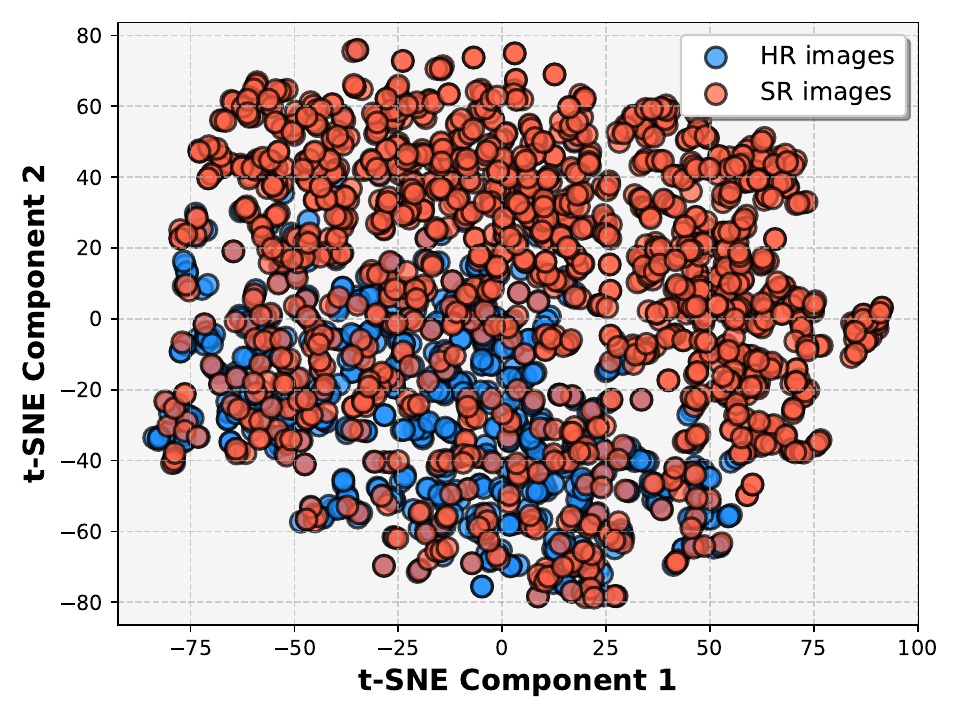}  & 
        \includegraphics[width=0.5\linewidth]{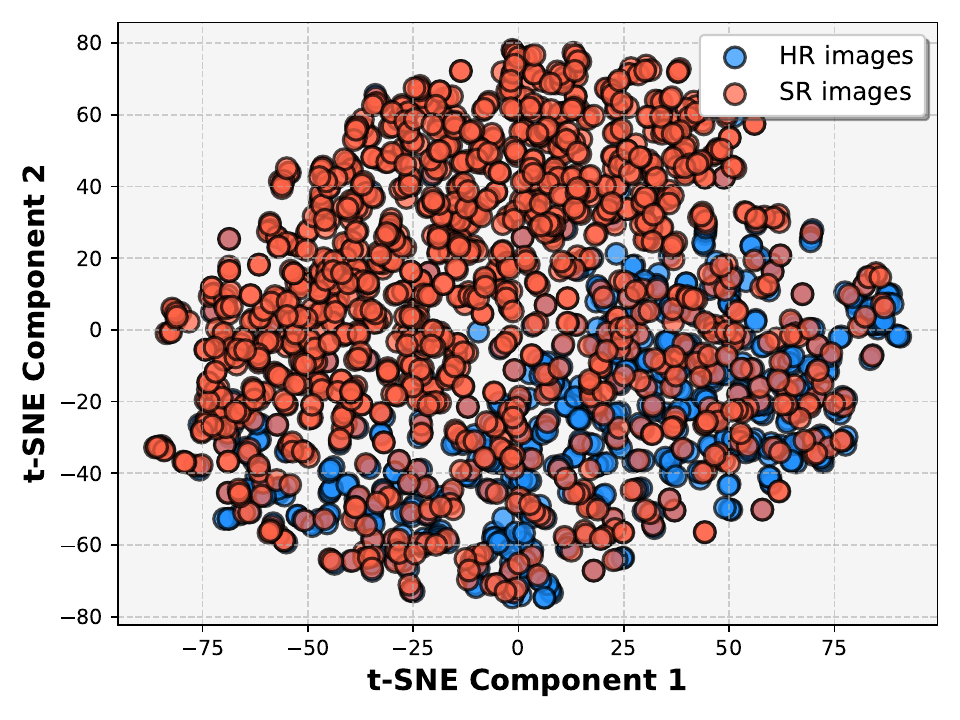}\\  
    \end{tabular}
    \caption{t-SNE visualization of feature embeddings for high-resolution and super-resolved images, with respect to the embedding distance on the CCPD dataset~\citep{xu2018towards}. (Left) Euclidean distance and (right) Manhattan distance.}
    \label{fig:tsne_ccpd}
\end{figure}

\begin{figure}[htbp]
    \centering
    \begin{tabular}{cc}
       \includegraphics[width=0.5\linewidth]{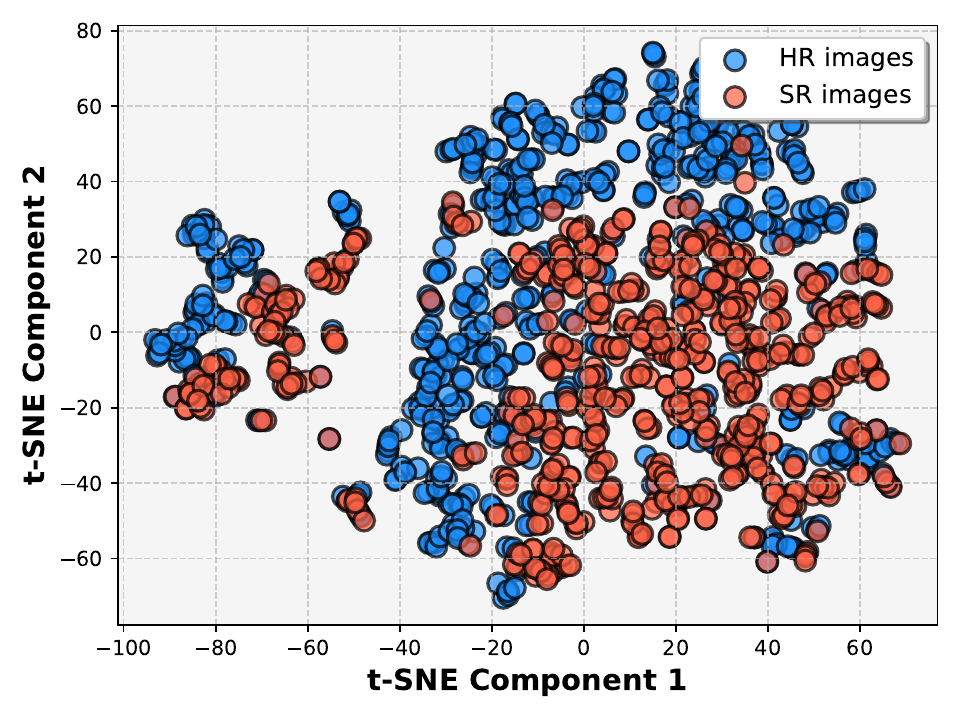}  & 
        \includegraphics[width=0.5\linewidth]{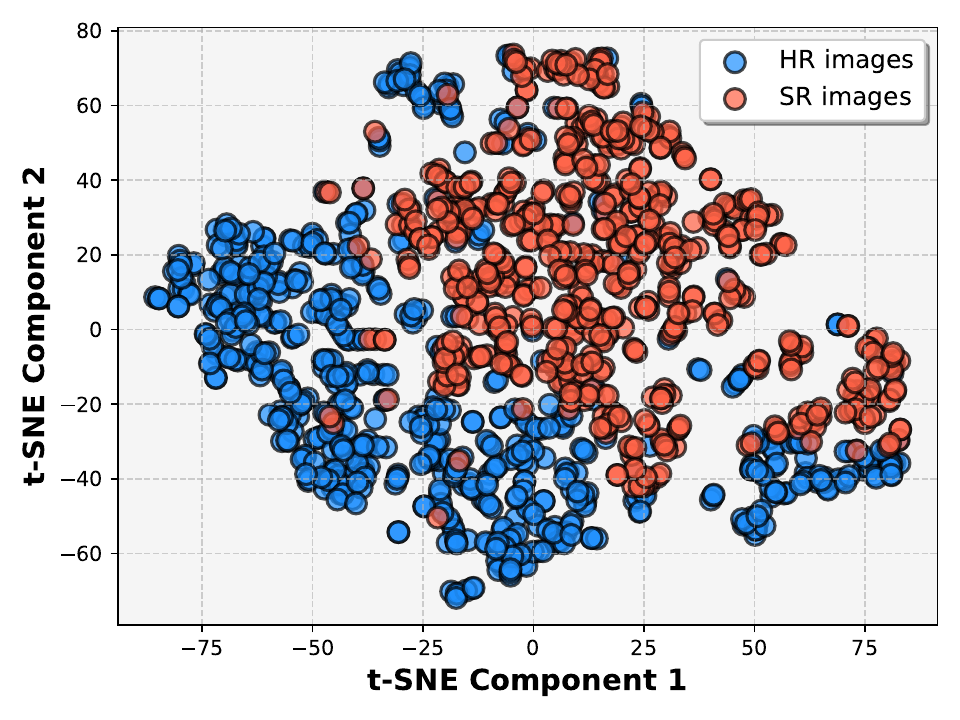}\\  
    \end{tabular}
    \caption{t-SNE visualization of feature embeddings for high-resolution and super-resolved images, with respect to the embedding distance on the PKU dataset~\citep{7752971}. (Left) Euclidean distance and (right) Manhattan distance.}
    \label{fig:tsne_pku}
\end{figure}

\textcolor{black}{
The t-SNE visualizations of feature embeddings on the CCPD dataset, shown in Fig.\ref{fig:tsne_ccpd}, indicate that both distance metrics produce compact and well-clustered embeddings in the latent space. Notably, Manhattan distance results in slightly tighter clustering compared to Euclidean distance, suggesting that it enforces a stronger alignment between HR and SR embeddings. However, the visualization on the PKU dataset, shown in Fig.\ref{fig:tsne_pku}, exhibits a different pattern. The embeddings appear more distinct, with well-separated clusters likely influenced by the varying characteristics of license plate images, particularly differences in background color. This distribution suggests that factors beyond the choice of distance metric contribute to the structure of the latent space in this dataset. Besides, the embedding distribution on the PKU dataset highlights it challenging nature, yet the model is able to generate SR images with similar embeddings to the HR one for several data points. 
}

\textcolor{black}{\subsubsection{Applicability to $\times4$ Super-Resolution}}

\textcolor{black}{While the primary focus of this work is on extreme $\times8$ super-resolution of license plate images, we also include an additional comparison at $\times4$ to align with recent literature and assess the generalization ability of the proposed method.}

\textcolor{black}{We adapted our model to operate under a $\times4$ upscaling setting and compared its performance to four recent and competitive methods: SwinIR~\citep{liang2021swinir}, DiT-SR~\citep{ditsr}, DiffTSR~\citep{zhang2024diffusion}, and SeeSR~\citep{wu2024seesr}. Performance is studied on the CCPD dataset~\citep{xu2018towards} because of its large size and the availability of labels. These models were selected based on their strong relevance and performance in text and general image super-resolution tasks.}

\textcolor{black}{Quantitative results are presented in Table~\ref{tab:x4comparison_qu}, focusing on visual fidelity (PSNR, SSIM) and perceptual similarity (LPIPS). Despite being optimized for the more challenging $\times8$ scale, our model demonstrates strong performance at $\times4$, indicating robust generalization. Among the compared methods, SeeSR performs the weakest, showing the lowest PSNR and SSIM scores and the highest LPIPS, which reflects poor reconstruction and weak perceptual quality. DiT-SR performs better but still lags behind the other approaches. SwinIR offers notable improvements in both PSNR and SSIM, along with relatively low LPIPS, highlighting its strength in preserving both details and structure. DiffTSR further improves upon SwinIR, achieving the second-best results across all metrics, particularly with a lower LPIPS score (0.0581). Our proposed model outperforms all competitors, achieving the highest PSNR (31.45), SSIM (0.8590), and the lowest LPIPS (0.0436), demonstrating its effectiveness in producing visually accurate and perceptually faithful images under moderate upscaling.}

\textcolor{black}{Beyond visual quality, Table~\ref{tab:x4comparison_rec} evaluates the impact of super-resolution on license plate recognition. The results reveal consistent trends. SeeSR again performs the worst, with poor accuracy and high error rates, indicating difficulty in preserving readable text. DiT-SR shows moderate improvements but remains below SwinIR and DiffTSR. SwinIR delivers strong recognition results, particularly in exact match accuracy and character-level metrics, confirming its reliability. DiffTSR improves further, ranking second in most recognition metrics, including word error rate and exact match.}

\textcolor{black}{Our model achieves the best performance overall, with the highest exact match accuracy (69.60\%), the lowest character error rate (9.80\%), and the top scores in precision, recall, and F1-score. These results highlight the model’s ability to restore sharp and readable text details, which is crucial for accurate license plate recognition. In summary, even though it is primarily designed for $\times8$ super-resolution, our method proves highly effective and competitive at $\times4$, both in terms of visual quality and practical downstream recognition.}

\begin{table*}[ht]
\centering
\rowcolors{1}{gray!10}{white}
\caption{Performance comparison of the proposed model with recent SR methods on the CCPD dataset~\citep{xu2018towards} under $\times4$. The median (\textcolor{gray}{$\pm$ standard deviation}) over the testing set is reported. The best and second-best performances are respectively highlighted in \textcolor{red}{\textbf{bold red}} and \textcolor{blue}{\textbf{bold blue}}.}
\label{tab:x4comparison_qu}
\begin{tabular}{lccc}
\toprule
\textbf{Method} & \textbf{PSNR} $\uparrow$ & \textbf{SSIM} $\uparrow$ & \textbf{LPIPS} $\downarrow$  \\
\midrule
SwinIR~\citep{liang2021swinir} & 30.94 (\textcolor{gray}{$\pm$ 1.29}) & 0.8396 (\textcolor{gray}{$\pm$ 0.094}) & 0.0792 (\textcolor{gray}{$\pm$ 0.027})  \\
DiT-SR~\citep{ditsr} & 29.73 (\textcolor{gray}{$\pm$ 0.94}) & 0.5825 (\textcolor{gray}{$\pm$ 0.063}) & 0.1546 (\textcolor{gray}{$\pm$ 0.048}) \\
SeeSR~\citep{wu2024seesr} & 28.26 (\textcolor{gray}{$\pm$ 0.57}) & 0.5251 (\textcolor{gray}{$\pm$ 0.092}) & 0.2032 (\textcolor{gray}{$\pm$ 0.080}) \\
DiffTSR~\citep{zhang2024diffusion} &  \textcolor{blue}{\textbf{31.26}} (\textcolor{gray}{$\pm$ 1.30}) & \textcolor{blue}{\textbf{0.8412}} (\textcolor{gray}{$\pm$ 0.082}) &  \textcolor{blue}{\textbf{0.0581}} (\textcolor{gray}{$\pm$ 0.010}) \\
\textbf{Ours (×4)} & \textcolor{red}{\textbf{31.45}} (\textcolor{gray}{$\pm$ 1.54}) & \textcolor{red}{\textbf{0.8590}} (\textcolor{gray}{$\pm$ 0.102}) & \textcolor{red}{\textbf{0.0436}} (\textcolor{gray}{$\pm$ 0.015})  \\
\bottomrule
\end{tabular}
\end{table*}

\begin{table*}[ht]
\centering
\rowcolors{1}{gray!10}{white}
\caption{Text recognition performance on super-resolved license plates from the CCPD dataset~\citep{xu2018towards} under $\times4$. The best and second-best results are highlighted in \textcolor{red}{\textbf{bold red}} and \textcolor{blue}{\textbf{bold blue}}, respectively.}
\label{tab:x4comparison_rec}
\begin{tabular}{l|l|l|l|l|l|l|l}
\toprule
\textbf{Method} & \textbf{EMA$\uparrow$} & \textbf{L-sim$\uparrow$} & \textbf{CER$\downarrow$} & \textbf{WER$\downarrow$} & \textbf{Prec.$\uparrow$} & \textbf{Recall$\uparrow$} & \textbf{F1-S$\uparrow$} \\
\midrule
SwinIR~\citep{liang2021swinir} & 67.10\% & 88.82\% & \textcolor{blue}{\textbf{11.18\%}} & 32.90\% & \textcolor{blue}{\textbf{83.27\%}} & \textcolor{blue}{\textbf{83.35\%}} & \textcolor{blue}{\textbf{83.21\%}} \\
DiT-SR~\citep{ditsr} & 56.80\% & 84.57\% & 15.43\% & 43.20\% & 76.66\% & 76.75\% & 76.57\% \\
SeeSR~\citep{wu2024seesr} & 28.99\% & 49.14\% & 59.86\% & 60.01\% & 42.23\% & 42.35\% & 41.10\% \\
DiffTSR~\citep{zhang2024diffusion} & \textcolor{blue}{\textbf{69.09\%}} & \textcolor{blue}{\textbf{89.95\%}} & 12.05\% & \textcolor{blue}{\textbf{31.80\%}} & 82.12\% & 82.10\% & 82.00\%\\
\textbf{Ours (×4)} & \textcolor{red}{\textbf{69.60\%}} & \textcolor{red}{\textbf{90.20\%}} & \textcolor{red}{\textbf{9.80\%}} & \textcolor{red}{\textbf{30.40\%}} & \textcolor{red}{\textbf{85.52\%}} & \textcolor{red}{\textbf{85.54\%}} & \textcolor{red}{\textbf{85.45\%}}  \\
\bottomrule
\end{tabular}
\end{table*}

\section{Conclusion}

\textcolor{black}{
This article presents a novel framework for license plate super-resolution (LPSR) that substantially outperforms existing methods by integrating a pixel and embedding consistency loss (PECL). The proposed approach, which leverages a Siamese network with contrastive loss, effectively balances pixel-level fidelity with embedding-level similarity, leading to consistent improvements in PSNR, SSIM, LPIPS, and OCR metrics. Experimental results demonstrate that our method not only recovers fine details and reduces blurring and compression artifacts but also enhances semantic consistency, as evidenced by superior text recognition performance on the CCPD dataset, achieving the highest exact match accuracy and the lowest error rates among competing methods.}

\textcolor{black}{
Furthermore, qualitative comparisons across datasets reveal that while methods such as ESRGAN perform relatively well on the smaller-scale PKU dataset, our model consistently delivers more natural textures and sharper edges, thereby ensuring robust performance under diverse degradation conditions. The dynamic optimization of pixel-wise and embedding-based losses during training ensures stability and adaptability, ultimately yielding an SR output that is both visually accurate and task-aware. In summary, this work advances LPSR by combining an innovative loss function with a robust architectural design, demonstrating significant improvements over state-of-the-art techniques.} 

\textcolor{black}{
Future work will explore multi-image super-resolution to leverage temporal and spatial information for enhanced reconstruction. Besides, the proposed model may benefit from knowledge distillation to create a lightweight and more robust version.}

\section{Akgnowledement}
This work is supported by The French Research Funding Agency (ANR) under project IMPROVED ANR-22-CE39-0006.

\bibliographystyle{cas-model2-names}
\bibliography{refs}

\end{document}